%% file: paper.tex
\documentclass[10pt,a4paper]{article}
\usepackage{float}
\usepackage{multirow}
\usepackage{graphics}
\usepackage[cp1252,utf8]{inputenc}
\usepackage{hyperref}
\usepackage{graphicx} 
\usepackage{amsmath}
\usepackage{amsfonts}
\usepackage{amssymb}

\usepackage[top=1in, bottom=2cm, left=2cm, right=2cm]{geometry}

\usepackage{authblk}

\usepackage{changepage} 

\usepackage{afterpage}

\usepackage{placeins}

\usepackage{wrapfig}

\title{\bf Shadow of a noncommutative geometry inspired\\ Ay$\acute{o}$n Beato\@ Garc$\acute{i}$a black hole}

\author[a]{Ashis Saha \thanks{sahaashis0007@gmail.com, ashisphys18@klyuniv.ac.in}}
\author[b]{Sai Madhav Modumudi \thanks{saimadhav.modumudi@gmail.com}}
\author[c]{Sunandan Gangopadhyay \thanks{sunandan.gangopadhyay@gmail.com, sunandan@associates.iucaa.in}}

\affil[a]{\footnotesize Department of Physics, University of Kalyani, Kalyani, India}
\affil[b]{\footnotesize Department of Physical Sciences, Indian Institute of Science Education and Research - Kolkata, India}
\affil[c]{\footnotesize Department of Theoretical Sciences, S.~N.~Bose National Centre for Basic Sciences, JD Block, Sector- III, Salt Lake, Kolkata 700106, India}

\date{}

\begin{document}

\maketitle

\begin{abstract}
\noindent We introduce the noncommutative geometry inspired Ay$\acute{o}$n Beato\@ Garc$\acute{i}$a black hole metric and study various properties of this metric by which we try to probe the allowed values of the noncommutative parameter $\vartheta$ under certain conditions. We then construct the shadow (apparent shape) cast by this black hole. We derive the corresponding photon orbits and explore the effects of noncommutative spacetime on them. We then study the effects of noncommutative parameter $\vartheta$, smeared mass $m(r)$, smeared charge $q(r)$ on the silhouette of the shadow analytically and present the results graphically. We then discuss the deformation which arises in the shape of the shadow under various conditions. Finally, we introduce a plasma background and observe how the shadow behaves in this scenario.\\
\end{abstract}

\section{Introduction}

\noindent From the observational point of view, the invisibility of black hole has been a matter of concern from the very birth of a geometrical theory of gravity. The physical sense of observation implies that one has to interact with the geometrical structure of the black hole by which information is gained; in this case, photon or its dual nature ``light" does the work. So for the last few decades people are trying to visualize black holes in order to get insight about its geometrical features. The well known application of general relativity (GR) known as the gravitational lensing has been a strong candidate to study structures of various massive sources of gravity. It has been observed that when the gravitational field of the concerned object becomes very strong, the results become very interesting. This occurs due to the various classifications of the photon orbits. The shadow of a black hole can be realized as the result of strong gravitational lensing creating a dark spot in the sky of a distant observer, or more precisely it is the optical appearance displayed by the black hole when there is a bright distant source behind it, and there is a uniform distribution of photons except for the region between the black hole and the infinitely distant observer from where optical information is obtained. In \cite{Synge}, it was shown that for a non-rotating black hole the shadow appears to be a perfect circle and later in \cite{Bardeen}, it was found that the shadow of a Kerr black hole is not circular anymore. The shape was further deformed when charge came into play \cite{Kerr-Newmann Shadow}. The angular radius of the Schwarzschild black hole was first calculated  in \cite{Synge}, after that in \cite{Hioki}, two physical observables were introduced in the mathematical calculations of the shadow. Thereafter, studies have been carried out for various types of black holes in \cite{Ghosh}-\cite{5DBH}. The event horizon telescope \cite{EHT} is an important project underway which would give us valuable information about the nature of black holes.\\
The occurrence of curvature singularity in the black hole spacetime is one of the most challenging and fascinating problems in GR. Noncommutative (NC) geometry is a promising candidate to resolve this issue since it probes physics at very small length scales. It basically implies that it is not possible to localize a particle accurately at very small length scales. In NC geometry, the notion of a point mass gets replaced by a smeared mass at short distances \cite{Nicolini},\cite{SG}. Further, in this framework, the coordinates of spacetime satisfies the relation \cite{Nicolini}
$$[x^{i},x^{j}] =i\vartheta^{ij}$$
 where $\vartheta$ is a $D\times D$ ($D$ denotes the dimension of spacetime) real-valued, antisymmetric matrix.\\
As we have stated earlier, the notion of a point mass is replaced by a smeared mass or a Gaussian distribution of mass-density $${\rho_{\vartheta}(r)=\frac{M}{(4\pi\vartheta)^{3/2}}\exp(-r^2/4\vartheta)}$$ where $\sqrt{\vartheta}$ has the dimension of length and corresponds to the parameter appearing in the noncommutating coordinates \cite{SG 2}. Maggiore \cite{Maggiore} suggested that this new short-length scale occurs naturally from any physically possible quantum theory of gravity.\\
In this article, the main objective is to investigate the shadow of a noncommutative geometry (NC) inspired Ay$\acute{o}$n Beato\@ Garc$\acute{i}$a (ABG) black hole. The motivation of carrying out this investigation is to study the effect of spacetime noncommutativity on the shadow cast by a black hole. This would give us important information about the structure of spacetime at very small length scales which can in principle be tested. A study of the NC Kerr-Newman black hole was carried out in \cite{epjc}. \\
\noindent The paper is organized as follows. In section \ref{sec2}, we have introduced noncommutativity inspired ABG metric and obtained the null geodesics. We then obtained the effective potential and made a brief discussion on the unstable circular orbits. In section \ref{sec3}, we then studied the celestial coordinates with the help of null geodesics and graphically constructed the shadow of the black hole. In section \ref{sec4}, we give a brief review on the observable parameters of the black hole shadow and compute them. In section \ref{sec5}, we have introduced the plasma background and obtained the new set of null geodesics which leads to new results for black hole shadows. We conclude in section \ref{sec6}.\\
\noindent In this paper, we shall set $G=c=\hbar=1$.

\section {Noncommutative inspired Ay$\acute{o}$n Beato\@ Garc$\acute{i}$a (ABG) black hole spacetime}\label{sec2}

\begin{figure}[!h]
	\begin{minipage}[t]{0.48\textwidth}
		\centering\includegraphics[width=\textwidth]{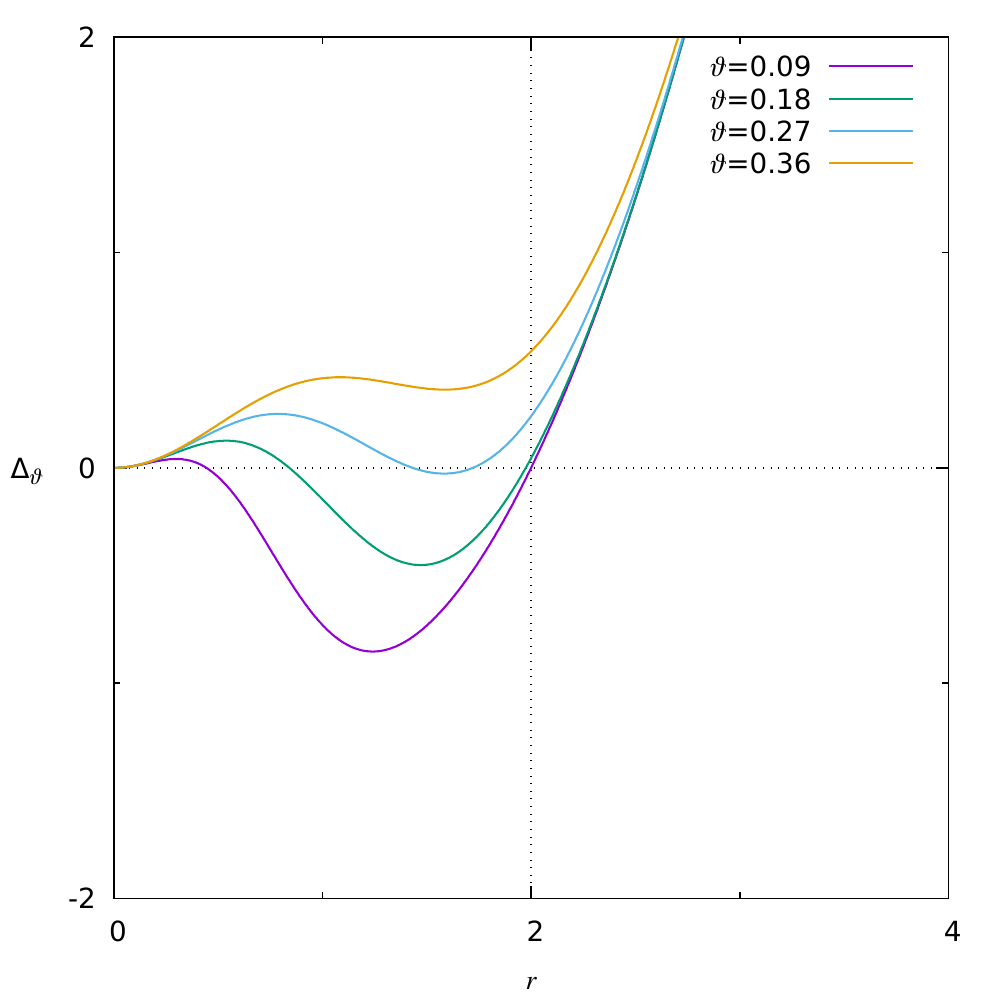}\\
		{\footnotesize $Q=0$, $a=0$}
	\end{minipage}\hfill
	\begin{minipage}[t]{0.48\textwidth}
		\centering\includegraphics[width=\textwidth]{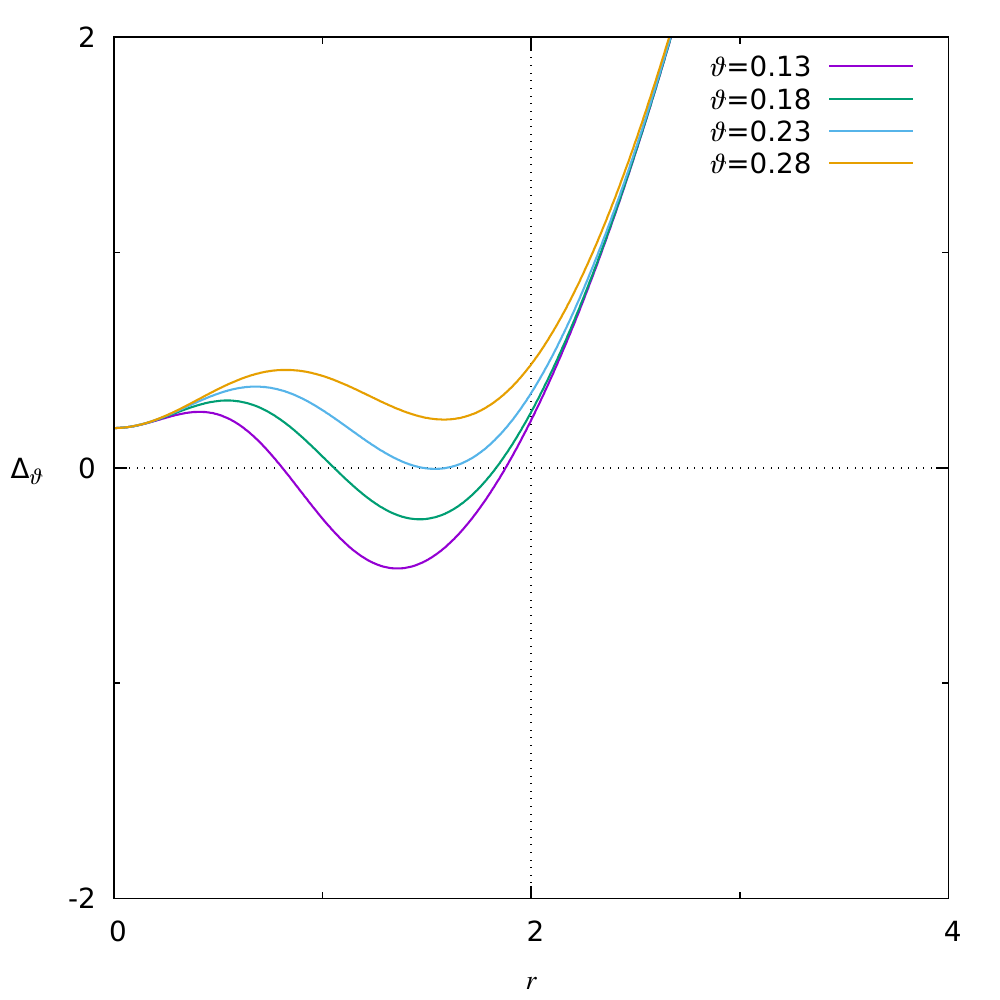}\\
		{\footnotesize $Q=0.11$, $a=0.43$}
	\end{minipage}
	\caption{Effect of NC parameter $\vartheta$ on $\Delta_{\vartheta}$}
	\label{ehori0}
\end{figure}

\begin{figure}[!h]
	\begin{minipage}[t]{0.48\textwidth}
		\centering\includegraphics[width=\textwidth]{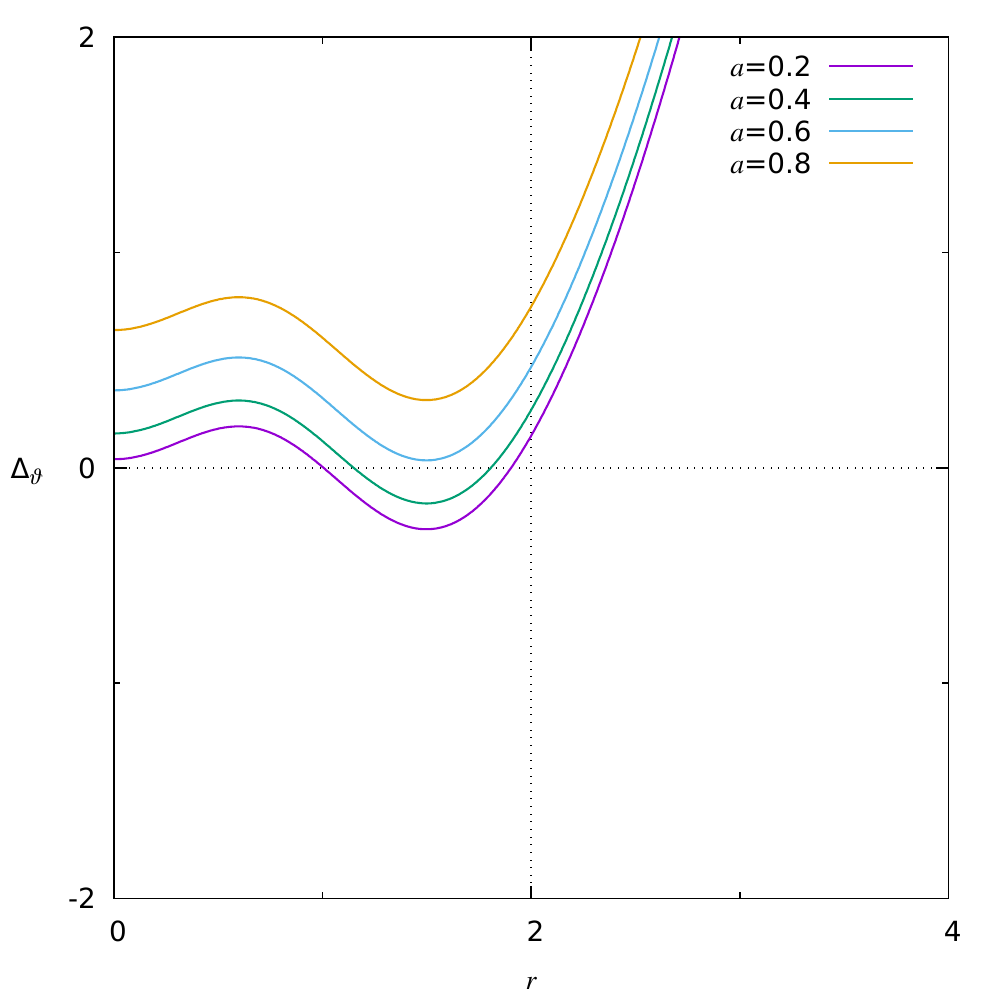}\\
		{\footnotesize $Q=0.12$, $\vartheta=0.2$}
	\end{minipage}\hfill
	\begin{minipage}[t]{0.48\textwidth}
		\centering\includegraphics[width=\textwidth]{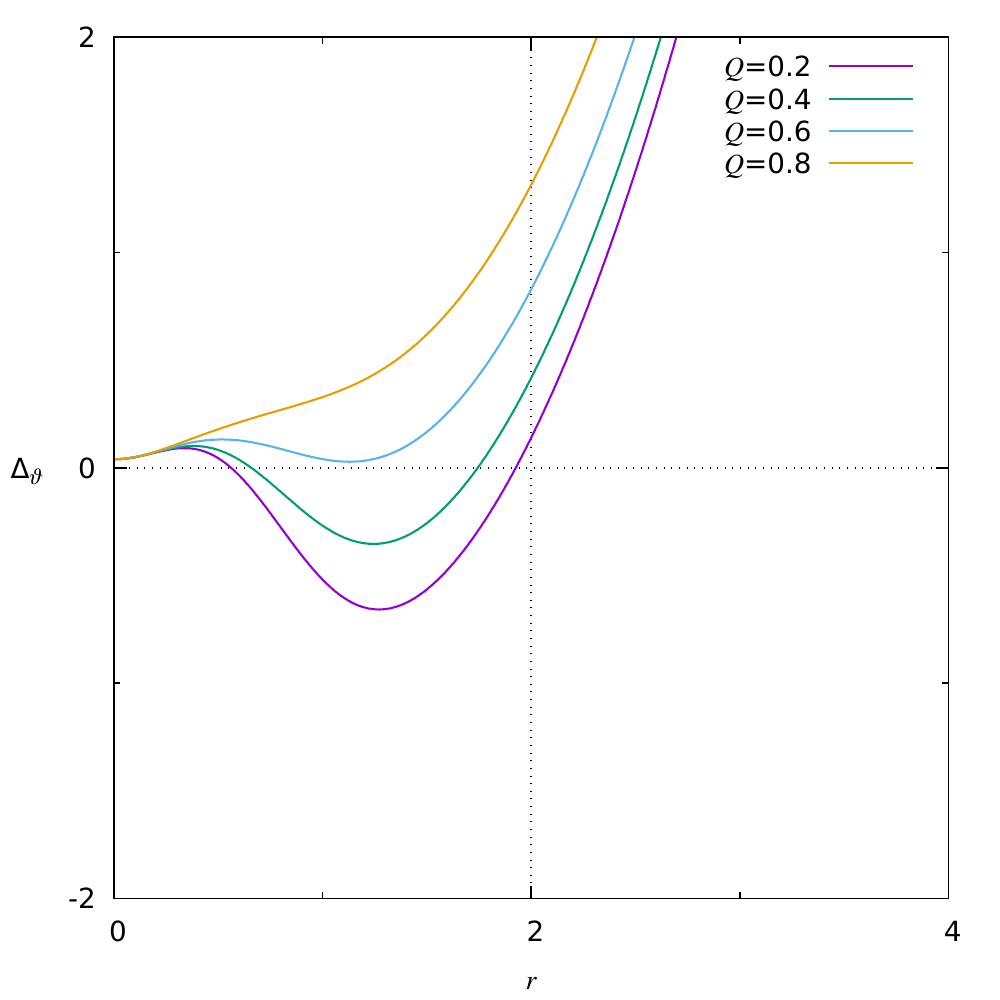}\\
		{\footnotesize $a=0.2$, $\vartheta=0.1$}
	\end{minipage}
	\caption{Effect of $a$ and $Q$ on $\Delta_{\vartheta}$}
	\label{ehori}
\end{figure}

The ABG spacetime was introduced in \cite{ABG} and it is an exact non-singular solution of the Einstein equation coupled with nonlinear electrodynamics. The rotating case of ABG was obtained in \cite{rotating abg} and \cite{rotating abg 2} based on the Newman-Janis algorithm \cite{newman-janis}. The metric of ABG black hole in Boyer-Lindquist coordinates reads
\begin{equation}
\begin{split}
ds^2\ =\ &-f(r,\theta)dt^2\ +\ \frac{\Sigma}{\Delta}dr^2\ +\ \Sigma d\theta^2\ +\ \sin^2\theta[\Sigma-a^2(f(r,\theta)-2)\sin^2\theta]d\phi^2\\[5pt] &-\  2a\sin^2\theta[1-f(r,\theta)]dtd\phi \qquad
\label{cmetric}
\end{split}
\end{equation}
where
\begin{eqnarray}
f(r,\theta)=1-\frac{2Mr\sqrt{\Sigma}}{(\Sigma+Q^2)^\frac{3}{2}} +\frac{Q^2\Sigma}{(\Sigma+Q^2)^2}~,\quad
\Delta =\Sigma f(r,\theta)+a^2\sin^2\theta~, \quad\Sigma=r^2+a^2\cos^2\theta
\end{eqnarray}
with $M$ being the mass, $a$ the rotation parameter and $Q$ the electric charge of the black hole.\\
\noindent In the realm of NC geometry, it is well known that a point-like gravitational source gets replaced by a smeared mass $m(r)$. The smearing of mass is introduced by a Gaussian mass density
\begin{equation}
{\rho_{\vartheta}(r)=\frac{M}{(4\pi\vartheta)^{3/2}}\exp(-r^2/4\vartheta)}
\end{equation}
together with a Gaussian charge density
\begin{equation}
{\rho_{e}(r)=\frac{Q}{(4\pi\vartheta)^{3/2}}\exp(-r^2/4\vartheta)}
\end{equation}
for the case of a charged black hole \cite{E spallucci}. This modifies the mass and the charge of the black hole to
\begin{equation}
\label{NCm}
 m(r) = {\frac{2M}{\sqrt{\pi}}\gamma(3/2,r^2/4\vartheta)}
\end{equation}
\begin{equation}
\label{NCq}
q^2(r) = \frac{Q^2}{\pi}\left[\gamma^2(1/2,r^2/4\vartheta)-\frac{r}{\sqrt{2\vartheta}}\gamma(1/2,r^2/2\vartheta)+r\sqrt{\frac{2}{\vartheta}}\gamma(3/2,r^2/4\vartheta)\right]
\end{equation}
where $\gamma(s,x)$ denotes the incomplete gamma function given by
\begin{equation}
\gamma\mathopen{}\left(s,x\right)\mathclose{} = \int_{0}^{x} t^{s-1} \mbox{e}^{-t}dt\ .
\end{equation}
Following the approach in \cite{newman-janis}, one finds the metric for the NC insiped ABG black hole to be
\begin{equation}\label{nc_abg_met} 
\begin{split}
ds^2\ =\ &-f_{\vartheta}(r,\theta)dt^2\ +\ \frac{\Sigma}{\Delta_{\vartheta}}dr^2\ +\ \Sigma d\theta^2\ +\ \sin^2\theta[\Sigma-a^2(f_{\vartheta}(r,\theta)-2)\sin^2\theta]d\phi^2\\-\ &2a\sin^2\theta[1-f_{\vartheta}(r,\theta)]dtd\phi \qquad
\end{split}
\end{equation}
where
\begin{eqnarray*}
f_{\vartheta}(r,\theta)=1-\frac{2m(r)r\sqrt{\Sigma}}{(\Sigma+q(r)^2)^\frac{3}{2}} +\frac{q(r)^2\Sigma}{(\Sigma+q(r)^2)^2}\\[10pt]
\Delta_{\vartheta}=\Sigma f_{\vartheta}(r,\theta)+a^2 \sin^2\theta
\qquad \Sigma=r^2+a^2\cos^2\theta~.
\end{eqnarray*}
The above form for the metric is obtained by replacing $M$ by $m(r)$ (\ref{NCm}) and $Q$ by $q(r)$ (\ref{NCq}). To construct the shadow of this black hole, we need to obtain the geodesic equations of photon for the NC inspired ABG metric. In order to obtain the geodesics of the photon, we use the Hamilton-Jacobi formalism. However, $f_{\vartheta}(r,\theta)$ has a complicated form which creates complications to work with. Therefore, we resolve this by making an approximation in $\theta$, as $\theta \approx \pi/2 + \epsilon$, where, $\epsilon$ is a small angle \cite{prd}. Physically this approximation implies that we consider nearly equitorial plane photon orbits, although it is not necessary that the unstable circular orbits of the photons will only be contained in this nearly equitorial regime. However, this fact does not take away the essence of the subsequent analysis made in this paper since the aim of our work is to study the effects of the spin $(a)$, charge $(Q)$ and NC parameter $(\vartheta)$ on the shadow of the  black hole spacetime considered in this paper. Further, in section \ref{sec3}, we have shown that the celestial coordinate system in which the shadow is constructed, indicates that for a observer far away from the black hole, the photons will reach the observer from regions near the equitorial plane. This simplification makes $\sin\theta \approx 1$ and $\cos\theta \approx \epsilon$ in the metric and simplifies $f_{\vartheta}(r,\theta)$ to $f_{\vartheta}(r)$
\begin{equation}
{f_{\vartheta}(r)=1-\frac{2m(r)r^2}{(r^2+q(r)^2)^\frac{3}{2}} +\frac{q(r)^2r^2}{(r^2+q(r)^2)^2}}\ .
\end{equation}
\noindent The event horizon of this black hole is given by the largest root of the equation $\Delta_{\vartheta} = 0$, which reads
\begin{equation}
r^2-\frac{2m(r)r^4}{(r^2+q(r)^2)^\frac{3}{2}} +\frac{q(r)^2r^4}{(r^2+q(r)^2)^2}+a^2=0\ .
\end{equation}
In figures \ref{ehori0} and \ref{ehori}, we have shown the variation of $\Delta_{\vartheta}$ for different values of $a$, $Q$ and $\vartheta$. In all the figures, $M=1$.\\
\noindent With the above results in hand, we first observe that for different values of $a$ and $Q$, there is a maximum value of $\vartheta$ for which the metric gives a black hole solution. More precisely, for specific values of $a$ and $Q$ there is an upper bound on the NC parameter $\vartheta$ upto which the metric gives a black hole solution. In table \ref{ehr} we have numerically presented this fact which shows that for the NC inspired ABG black hole spacetime, the NC parameter $\vartheta < 0.276$ when spin parameter $a\in[0,1.0]$ and charge $Q\in[0,1.0]$ (for M=1).\\

\begin{table}[!h]
	\input{nc.tex}
	\caption{Event horizon radius $R_{EH}$ for different values of $(a,Q,\vartheta)$}
	\label{ehr}
\end{table}

\subsection {Geodesics in NC inspired ABG spacetime}

\noindent We now proceed to investigate the shadow of the NC inspired ABG black hole spacetime. The first step is to find the particle orbits in this spacetime. To do this, the standard approach is to exploit the directions of symmetry. Let $k^\mu$ be the four vector along the direction of symmetry and $u^\mu = \frac{dx^\mu}{d\sigma}$ be the tangent vector to the path of the particle $x^\mu=x^\mu(\sigma)$ where $\sigma$ is some affine parameter. It can now be shown by using the Killing equation that the component of $k^\mu$ along $u^\mu$ denoted by $u_\mu k^\mu$ is constant if the path $x^\mu$ is a geodesic \cite{Derek}.\\
Considering $k^\mu$ to be the time like direction since the metric coefficients are independent of time $t$, gives
\begin{equation}
k^\mu u_\mu=constant.
\end{equation}
Setting $k^\mu = (1,0,0,0)$ as $k^\mu$ corresponds to time like direction yields 
\begin{equation}
k^0 u_0=u_0=-E=constant.
\end{equation}
The constant $E$ represents the relativistic energy per unit mass of the particle relative to a stationary observer at infinity.
\noindent Observing that the metric coefficients are also independent of $\phi$ tells that the $\phi$-direction is also a direction of symmetry.
In this case, setting $k^\mu = (0,0,0,1)$ we have
\begin{equation}
k^\mu u_\mu=u_3=L_z=constant. 
\end{equation}
The constant $ L_z$ denotes the angular momentum per unit mass of the particle relative to a stationary observer at infinity. Using these constants of motion, one can easily obtain the geodesic equations along the directions of symmetry. This can be done by noting that
\begin{eqnarray}
u^0=g^{0\nu} u_\nu=g^{00} u_0+g^{03} u_3=-g^{00} E+g^{03} L_z\\[10pt]
u^3=g^{3\nu} u_\nu=g^{33} u_3+g^{30} u_0=g^{33} L_z-g^{30} E.
\end{eqnarray}
Substituting the contravariant components (which can be easily obtained from the metric (\ref{nc_abg_met})), the geodesic equations read
\begin{eqnarray}
r^2 \dot{t}=a(L_z-aE)+ \frac{r^2+a^2}{\Delta_{\vartheta}}\left[(r^2+a^2)E-aL_z\right] \\[10pt]
r^2 \dot{\phi}=(L_z-aE)+ \frac{a}{\Delta_{\vartheta}}\left[(r^2+a^2)E-aL_z\right].
\end{eqnarray}
The remaining geodesic equations for the NC inspired ABG black hole spacetime can be determined by using the Hamilton-Jacobi (HJ) equation which reads
\begin{equation}
{\frac{\partial{S}}{\partial{\lambda}} = {-\frac{1}{2}}g^{\mu\nu}{\frac{\partial S}{\partial {x^\mu}}\frac{\partial S}{\partial {x^\nu}}}}=-\frac{1}{2}g^{\mu\nu}p_{\mu}p_{\nu}
\label{hjeqn}
\end{equation}
where $\lambda$ is the affine parameter and $S$ is the Jacobi action. The HJ equation can be solved by using a simple ansatz of the form
\begin{equation}
S={\frac{1}{2}}m_0^2\lambda - Et + L_z\phi+S_r(r)+S_{\epsilon}(\epsilon)
\end{equation}
where $S_r(r), S_{\epsilon}(\epsilon)$ are functions of $r$ and $\epsilon$. Inserting this in eq.(\ref{hjeqn}), it can be shown after some lengthy algebra that
\begin{eqnarray}
r^2 \dot{r} &=& \sqrt{\mathcal{R_\vartheta}} \label{rad_nc} \\[7pt]
r^2 \dot{\epsilon} &=& \sqrt{\Theta}
\end{eqnarray}
where
\begin{equation}
\mathcal{R_\vartheta}=\left[E(r^2+a^2)-aL_z\right]^2 - \Delta_{\vartheta}\left[\mathcal{K}+(L_z-aE)^2\right],\ \mbox{and}
\end{equation}
\begin{equation}
\Theta = \mathcal{K}.
\end{equation}
Here, $\mathcal{K}$ is said to be the Carter constant \cite{carter}. Eq.(\ref{rad_nc}) can be recast as
\begin{equation}
(r^2 \dot{r})^2 + V_{eff}=0 
\end{equation}
where
\begin{equation}
V_{eff}\equiv\mathcal{R_{\vartheta}}=\left[E(r^2+a^2)-aL_z\right]^2 - \Delta_{\vartheta}\left[\mathcal{K}+(L_z-aE)^2\right]~.
\end{equation}
This equation describes the radial motion of the particle in the effective potential $ V_{eff}$.\\ The conditions for unstable circular orbits reads $V_{eff}=0$ and $\frac{\partial{V_{eff}}}{\partial{r}} = 0$. This leads to the following conditions
\begin{eqnarray}
(r^2+a^2-a\xi)^2 -\left[\eta+(\xi-a)^2\right]\Delta_{\vartheta}=0\\[10pt]
4r(r^2+a^2-a\xi)-\left[\eta+(\xi-a)^2\right]\Delta_{\vartheta}^\prime =0
\end{eqnarray}
where  $\xi = {L_z}/{E}$ and $\eta = \mathcal{K}/E^2$ are known as the Chandrasekhar constants \cite{chandrashekhar}.
Solving these equations, we obtain the expressions for $\xi$ and $\eta$ to be
\begin{eqnarray}
\xi &=& \frac{r^2+a^2}{a}- \frac{4r\Delta_{\vartheta}}{a\Delta_{\vartheta}^\prime} \label{xi_nc}\\[7pt]
\eta &=& \frac{4r(r^2+a^2-a\xi)}{\Delta_{\vartheta}^\prime}-(\xi-a)^2 \label{eta_nc}
\end{eqnarray}
where
$$\Delta_{\vartheta}^\prime=r^2 f_{\vartheta}^\prime(r) + 2rf_{\vartheta}(r)$$
and $\prime$ denotes derivative with respect to $r$. $\Delta_{\vartheta}^\prime$ contains the derivative of $f_{\vartheta}(r)$ with respect to $r$ and reads
\begin{multline}
f_{\vartheta}^\prime(r) = -\frac{2m^\prime(r)r^2}{(r^2+q(r)^2)^{3/2}}  - \frac{4m(r)r}{(r^2+q(r)^2)^{3/2}} + \frac{6m(r)r^2}{(r^2+q(r)^2)^{5/2}}(r+q(r)q^\prime(r)) + \frac{2q(r)q^\prime(r)r^2}{(q(r)^2+r^2)^2}\\[10pt] + \frac{2q(r)^2r}{(q(r)^2+r^2)^2} - (q(r)q^\prime(r)+r)\frac{4q(r)^2r^2}{(q(r)^2+r^2)^3}
\end{multline}
where
\begin{eqnarray}
\label{nc_m_der}
m^\prime(r) = \frac{Mr^2}{2\vartheta\sqrt{\pi\vartheta}}\exp\left(-r^2/ 4\vartheta\right)~, \quad
q^\prime(r) = \frac{Q^2}{2\pi q(r)}\mathcal{B}(r)
\label{nc_q_der}
\end{eqnarray}
and 
\begin{multline*}
\mathcal{B}(r)= \frac{2}{\sqrt{\vartheta}} \exp\mathopen{}\left(-\frac{r^2}{4\vartheta}\right)\mathclose{} \gamma\mathopen{}\left(\frac{1}{2},\frac{r^2}{4\vartheta}\right)\mathclose{}\  -\ \frac{r}{\vartheta} \exp\mathopen{}\left(-\frac{r^2}{4\vartheta}\right)\mathclose{}\  -\ \frac{1}{\sqrt{2\vartheta}}\ \gamma\mathopen{}\left(\frac{1}{2},\frac{r^2}{2\vartheta}\right)\mathclose{}\\[5pt]  +\ \frac{r^3}{2\sqrt{2}\vartheta^2} \exp\mathopen{}\left(-\frac{r^2}{4\vartheta}\right)\mathclose{}  +\ \sqrt{\frac{2}{\vartheta}}\ \gamma\mathopen{}\left(\frac{3}{2},\frac{r^2}{4\vartheta}\right)\mathclose{}~.
\end{multline*}
Substituting the above expressions in eqs.(\ref{xi_nc}) and (\ref{eta_nc}), we obtain 

\begin{eqnarray}
\xi = \frac{\mathcal{X}}{a\mathcal{D}} \label{xi_detail} \qquad
\eta = \frac{\mathcal{G}}{\mathcal{D}} \label{eta_detail}
\end{eqnarray}
where 
\begin{multline}
\mathcal{X}=(r^2+a^2)\left(-\frac{2m^\prime(r)r^4}{(r^2+q(r)^2)^{3/2}}  - \frac{4m(r)r^3}{(r^2+q(r)^2)^{3/2}} + \frac{6m(r)r^4}{(r^2+q(r)^2)^{5/2}}(r+q(r)q^\prime(r)) + \frac{2q(r)q^\prime(r)r^4}{(q(r)^2+r^2)^2}\right. \\[10pt] +\left. \frac{2q(r)^2r^3}{(q(r)^2+r^2)^2} - (q(r)q^\prime(r)+r)\frac{4q(r)^2r^4}{(q(r)^2+r^2)^3}+2r-\frac{4m(r)r^3}{(r^2+q(r)^2)^\frac{3}{2}} +\frac{2q(r)^2r^3}{(r^2+q(r)^2)^2}\right)-4r^3\\[5pt]
+\frac{8m(r)r^5}{(r^2+q(r)^2)^\frac{3}{2}} -\frac{4q(r)^2r^5}{(r^2+q(r)^2)^2}
\end{multline}

\begin{multline}
\mathcal{D}=-\frac{2m^\prime(r)r^4}{(r^2+q(r)^2)^{3/2}}  - \frac{4m(r)r^3}{(r^2+q(r)^2)^{3/2}} + \frac{6m(r)r^4}{(r^2+q(r)^2)^{5/2}}(r+q(r)q^\prime(r)) + \frac{2q(r)q^\prime(r)r^4}{(q(r)^2+r^2)^2}\\[10pt] + \frac{2q(r)^2r^3}{(q(r)^2+r^2)^2} - (q(r)q^\prime(r)+r)\frac{4q(r)^2r^4}{(q(r)^2+r^2)^3}+2r-\frac{4m(r)r^3}{(r^2+q(r)^2)^\frac{3}{2}} +\frac{2q(r)^2r^3}{(r^2+q(r)^2)^2} 
\end{multline}

\begin{multline}
\mathcal{G}=4r^3+4a^2r+2a\xi\left(-\frac{2m^\prime(r)r^4}{(r^2+q(r)^2)^{3/2}}  - \frac{4m(r)r^3}{(r^2+q(r)^2)^{3/2}} +\frac{6m(r)r^4}{(r^2+q(r)^2)^{5/2}}(r+q(r)q^\prime(r)) + \frac{2q(r)q^\prime(r)r^4}{(q(r)^2+r^2)^2} \right.\\[5pt]\left.+ \frac{2q(r)^2r^3}{(q(r)^2+r^2)^2} - (q(r)q^\prime(r)+r)\frac{4q(r)^2r^4}{(q(r)^2+r^2)^3}+2r-\frac{4m(r)r^3}{(r^2+q(r)^2)^\frac{3}{2}}+\frac{2q(r)^2r^3}{(r^2+q(r)^2)^2}-r\right)\\[5pt]
-(a^2+\xi^2)\left(-\frac{2m^\prime(r)r^4}{(r^2+q(r)^2)^{3/2}} - \frac{4m(r)r^3}{(r^2+q(r)^2)^{3/2}} + \frac{6m(r)r^4}{(r^2+q(r)^2)^{5/2}}(r+q(r)q^\prime(r)) + \frac{2q(r)q^\prime(r)r^4}{(q(r)^2+r^2)^2} \right.\\[5pt]\left. + \frac{2q(r)^2r^3}{(q(r)^2+r^2)^2} - (q(r)q^\prime(r)+r)\frac{4q(r)^2r^4}{(q(r)^2+r^2)^3}+2r-\frac{4m(r)r^3}{(r^2+q(r)^2)^\frac{3}{2}} +\frac{2q(r)^2r^3}{(r^2+q(r)^2)^2}\right)~.
\end{multline}

\vspace{0.2in}
\noindent In Table \ref{ehr}, we have computed the event horizon radius of the NC inspired ABG black hole for various combinations of the spin ($a$), charge ($Q$) and the NC parameter ($\vartheta$). The Table is later used to determine the allowed values of the NC parameter ($\vartheta$) for which the black hole shadow exists.
\section{Construction of Shadow}\label{sec3}

In this section, we shall study the shadow of the NC inspired ABG black hole with the geodesic equations and the conditions for unstable circular orbits in hand. To do this we are going to work in a new coordinate system ($\alpha,\beta $) known as the celestial coordinates \cite{celes.cord.}. These coordinates are defined as

\begin{wrapfigure}[17]{l}{0.55\textwidth}
	\centering
	\includegraphics[width=0.5\textwidth]{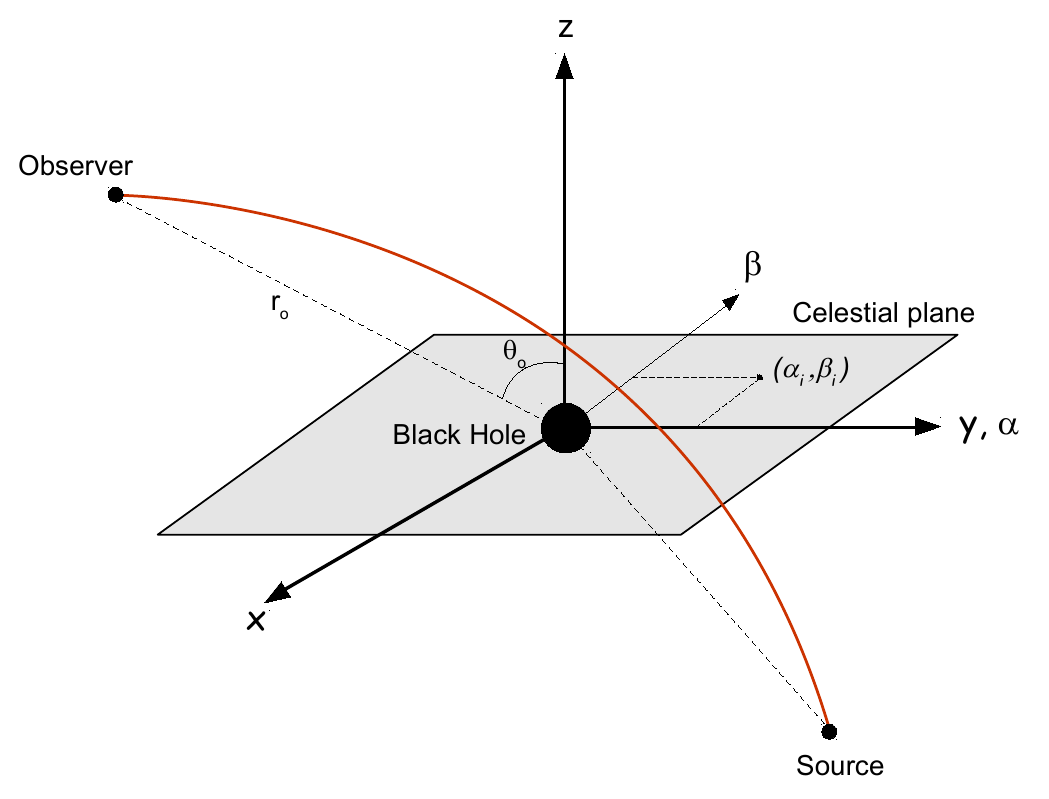}
	\caption{Celestial coordinates}
	\label{bs}
\end{wrapfigure}

$$\alpha=\lim_{r_0 \to \infty}\left(-r_0^2 \sin\theta_0 \frac{d\phi}{dr}\right)$$
$$\beta=\lim_{r_0 \to \infty}\left(r_0^2\frac{d\epsilon}{dr}\right).$$

\noindent The coordinate $\alpha$ denotes the apparent perpendicular distance of the shadow from the axis of symmetry, $\beta$ denotes the apparent perpendicular distance of the shadow from its projection on the equatorial plane. $r_0$ is the distance of the observer from the black hole, $\theta_0$ is the angle of inclination between the rotation axis of the black hole and observer's line of sight.

\noindent The values of $\frac{d\phi}{dr}$ and $\frac{d\epsilon}{dr}$ can be obtained by using the geodesic equations which are\\

\vspace{0.5in}

\begin{eqnarray}
\frac{d\phi}{dr}=\frac{(L_z-aE)+ \frac{a}{\Delta_{\vartheta}}[(r_0^2+a^2)E-aL_z]}{\sqrt{[E(r_0^2+a^2)-aL_z]^2 - \Delta_{\vartheta}[\mathcal{K}+(L_z-aE)^2]}}\\[10pt]
\frac{d\epsilon}{dr}=\frac{\sqrt{\mathcal{K}}}{\sqrt{[E(r_0^2+a^2)-aL_z]^2 - \Delta_{\vartheta}[\mathcal{K}+(L_z-aE)^2]}}~.
\end{eqnarray}

\begin{figure}[!h]
	\begin{minipage}[t]{0.48\textwidth}
		\centering\includegraphics[width=0.9\textwidth]{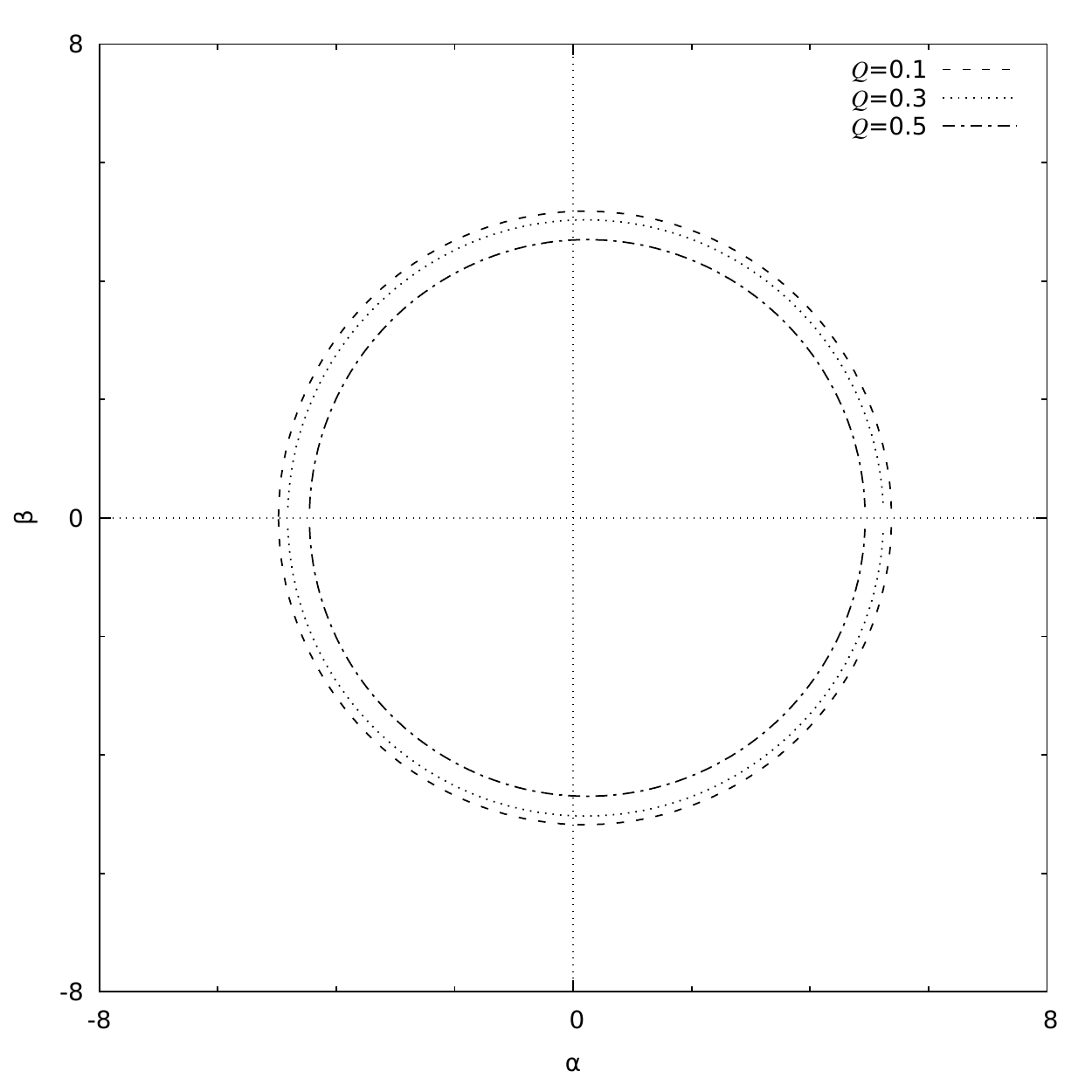}\\
		{\footnotesize $a=0.1$, $\vartheta=0.1$}
	\end{minipage}\hfill
	\begin{minipage}[t]{0.48\textwidth}
		\centering\includegraphics[width=0.9\textwidth]{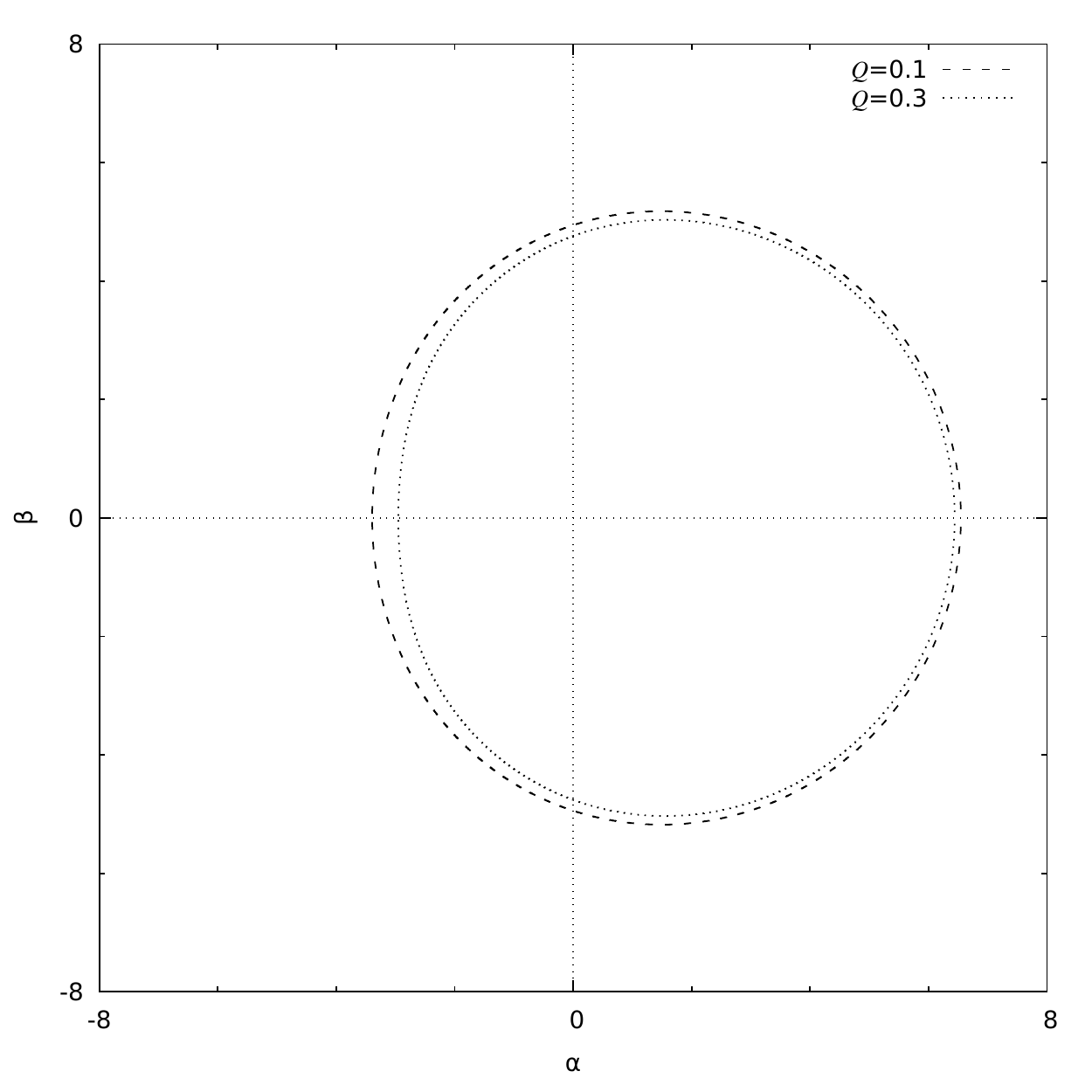}\\
		{\footnotesize $a=0.74$, $\vartheta=0.1$}
	\end{minipage}
	\caption{Varying $Q$ for different $a$ and $\vartheta$}
	\label{fig0}
\end{figure}

\begin{figure}[!h]
	\begin{minipage}[t]{0.48\textwidth}
		\centering\includegraphics[width=0.9\textwidth]{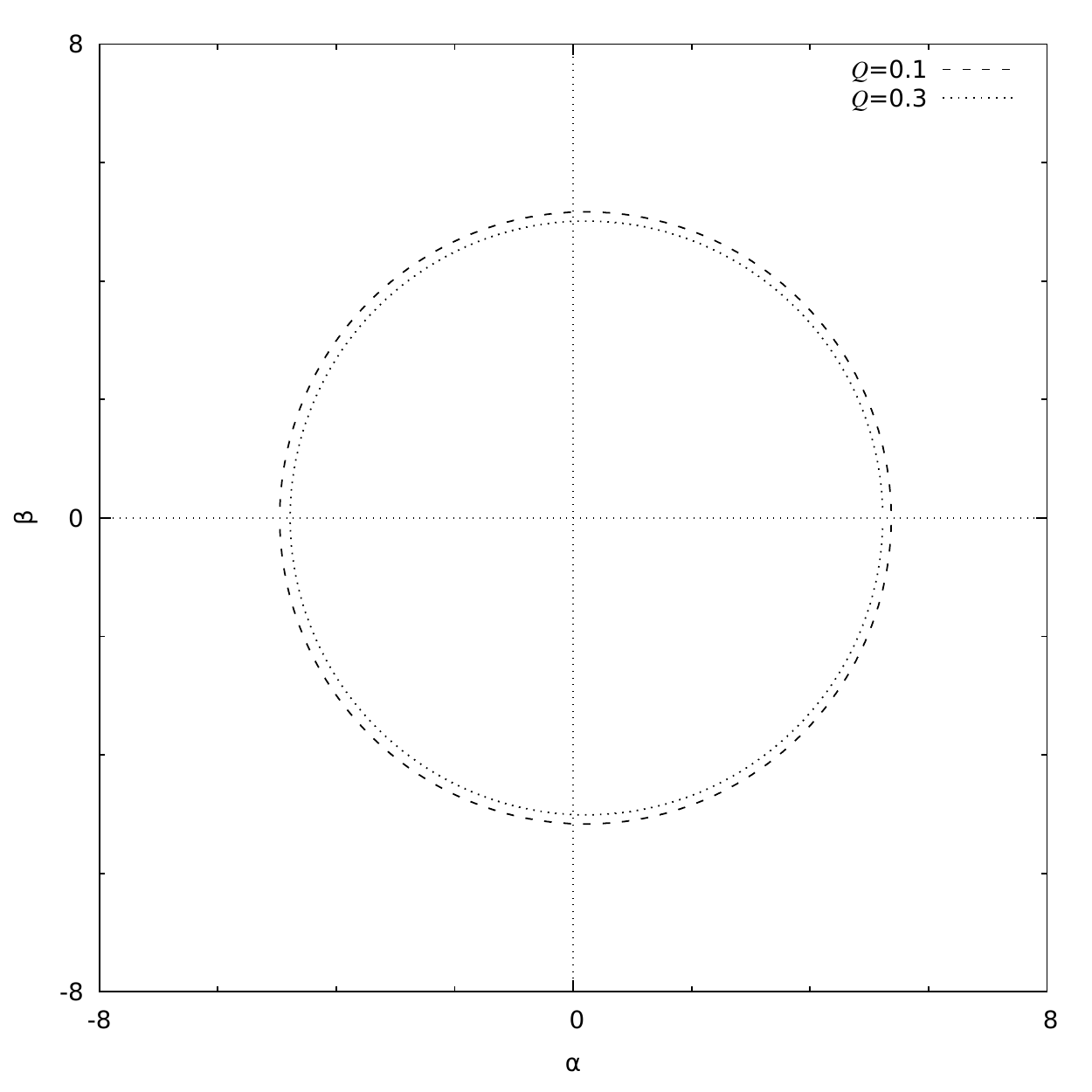}\\
		{\footnotesize $a=0.1$, $\vartheta=0.3$}
	\end{minipage}\hfill
	\begin{minipage}[t]{0.48\textwidth}
		\centering\includegraphics[width=0.9\textwidth]{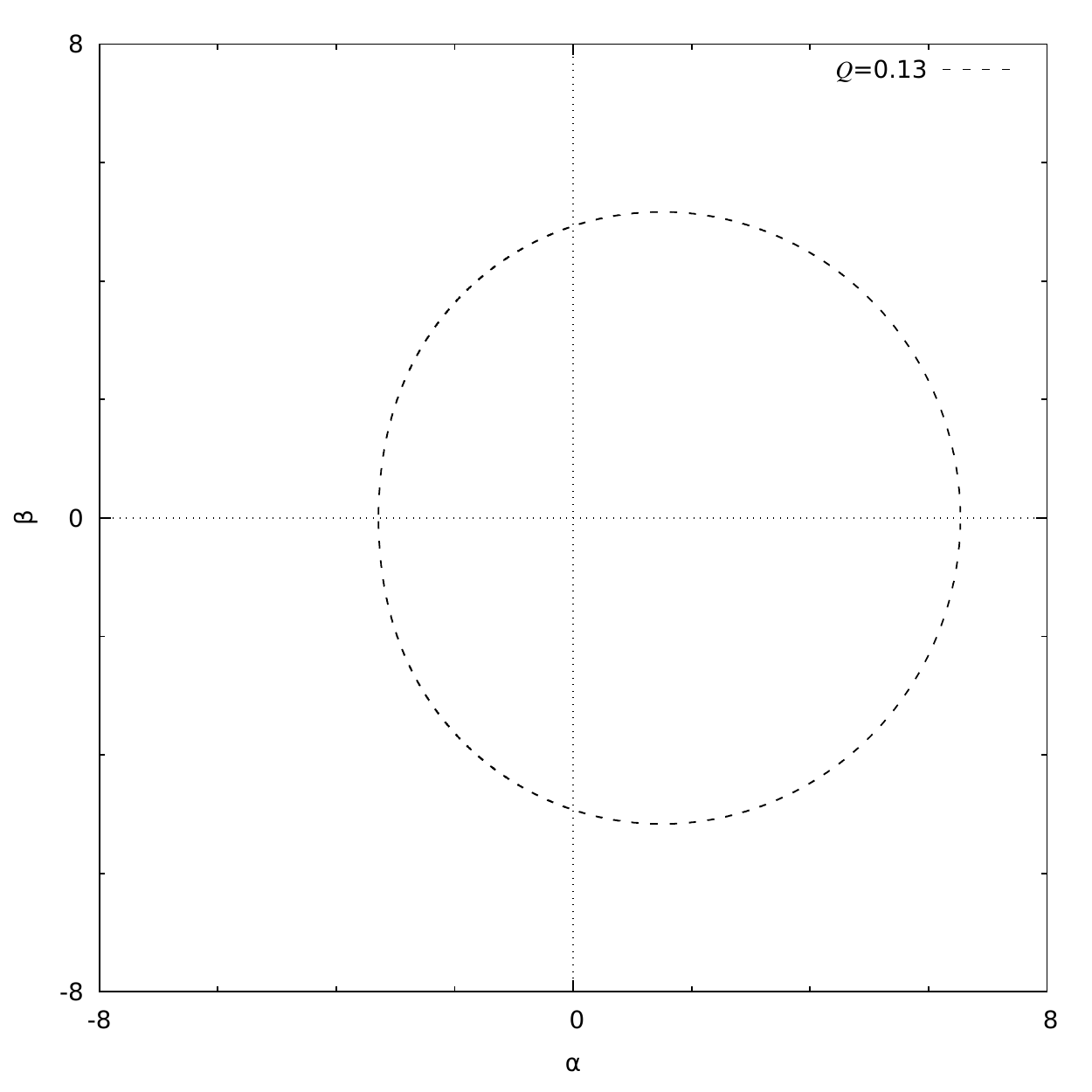}\\
		{\footnotesize $a=0.74$, $\vartheta=0.15$}
	\end{minipage}
	\caption{Varying $Q$ for different $a$ and $\vartheta$}
	\label{fig1}
\end{figure}

\begin{figure}[!h]
	\begin{minipage}[t]{0.48\textwidth}
		\centering\includegraphics[width=\textwidth]{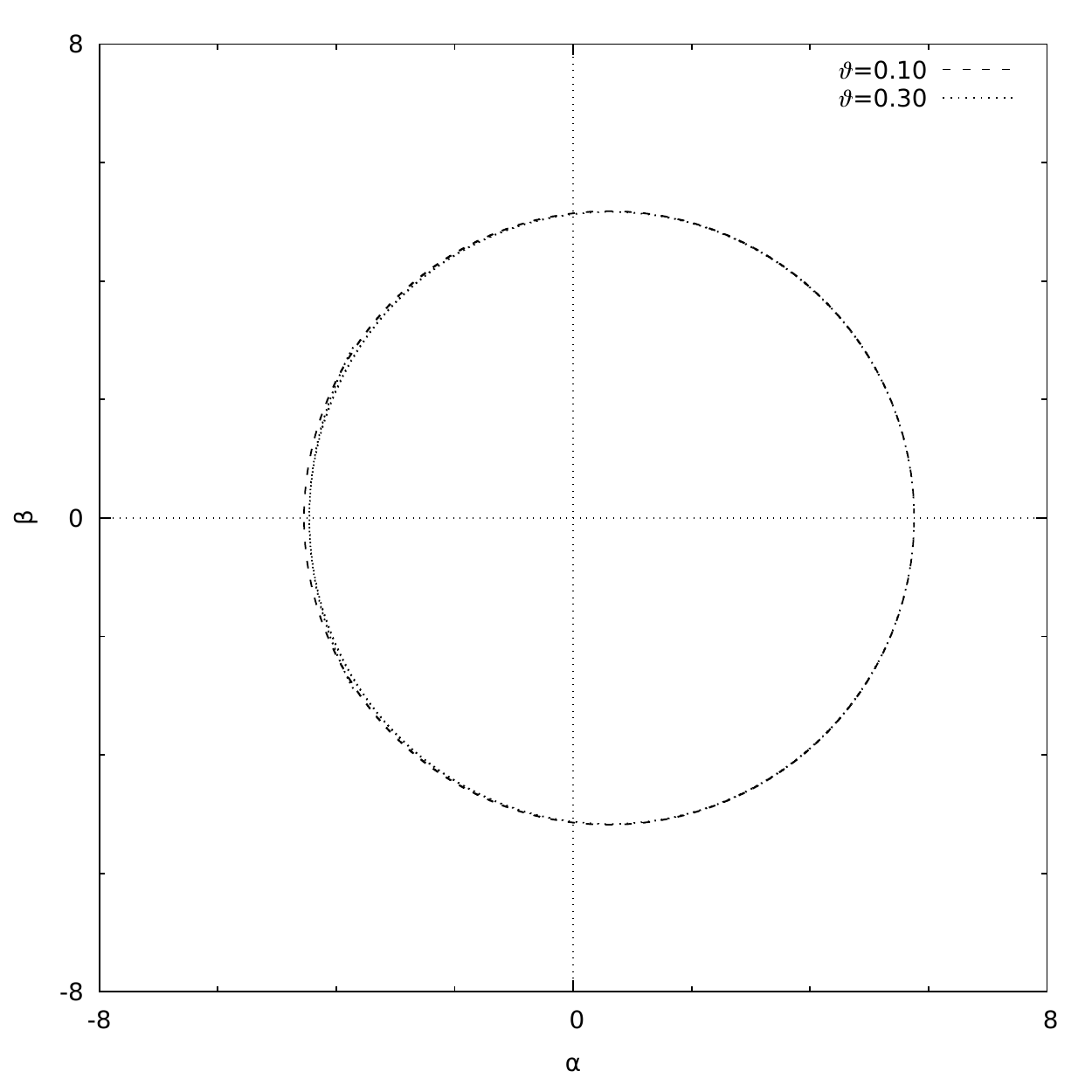}\\
		{\footnotesize $a=0.3$, $Q=0.1$}
	\end{minipage}\hfill
	\begin{minipage}[t]{0.48\textwidth}
		\centering\includegraphics[width=\textwidth]{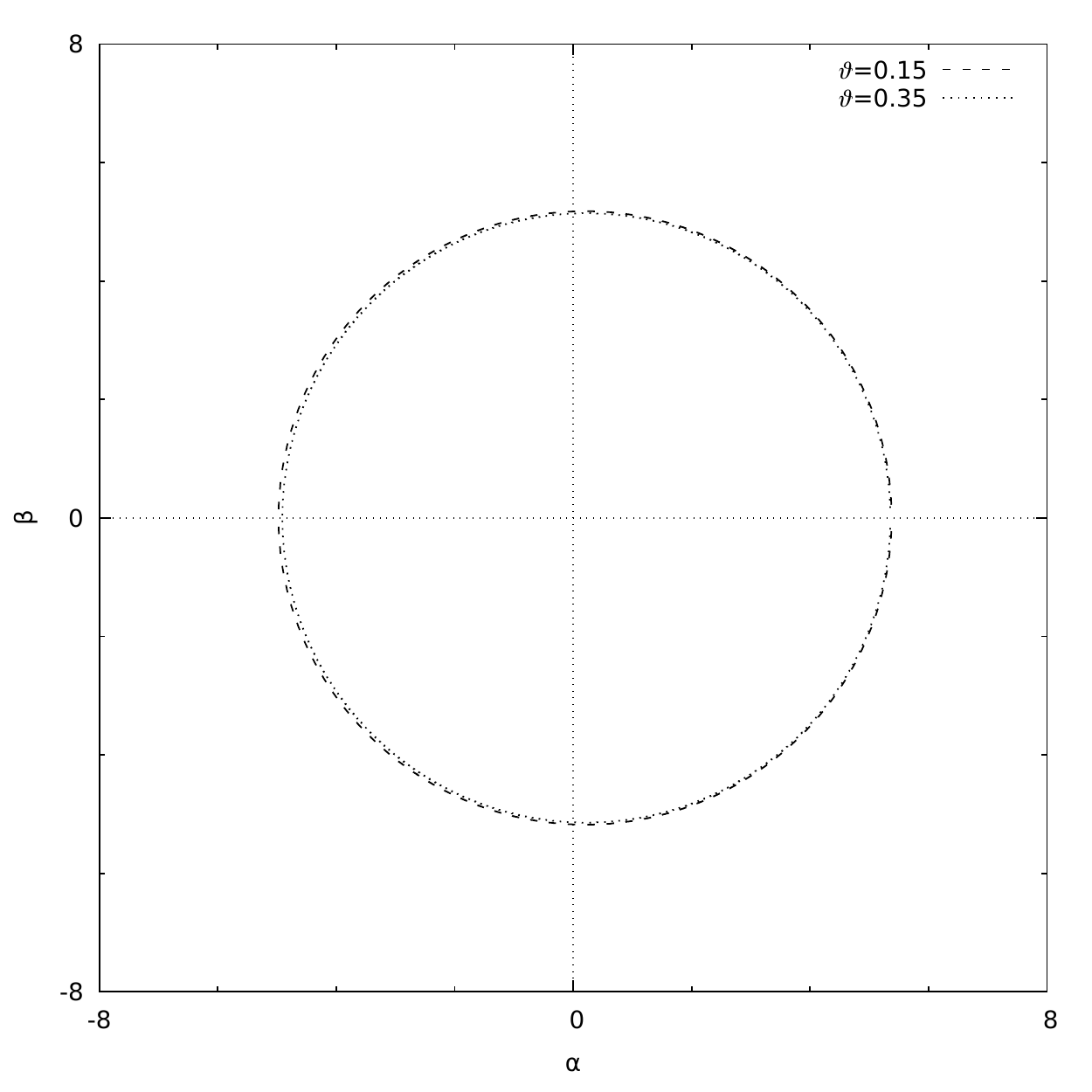}\\
		{\footnotesize $a=0.1$, $Q=0.1$}
	\end{minipage}
	\caption{Effect of $\vartheta$ on shadow of the black hole}
	\label{fig2}
\end{figure}

\noindent Substituting these values in the definitions of $ \alpha$ and $\beta$, and after taking the limit $r_0 \to \infty$ we get

\begin{equation}
\alpha = -\xi \qquad
\beta = \pm\sqrt{\eta}
\end{equation}
\noindent where $\xi$ and $\eta$ are given in eq.(\ref{xi_detail}) and (\ref{eta_detail}).\\
\noindent The shadows of the NC inspired ABG black hole metric can be obtained by plotting $\alpha$ vs $\beta$ for various values of rotation parameter $a$ and charge $Q$. These plots depict the shadows and also captures the effects of spacetime noncommutativity. In order to make the plots, we have to choose a value of spin ($a$) and charge ($Q$) for a fixed value of $\vartheta$. Since we are interested in the black hole solution of the spacetime, hence the choices cannot be made arbitrarily as it can lead to shadows of a naked singularity. The allowed values of $\vartheta$ for a fixed value of spin ($a$) and charge ($Q$) or any other allowed  combination is given in Table \ref{ehr}. The plots  constructed according to the allowed values given in Table \ref{ehr} leads to the shadows of the NC inspired ABG spacetime. These plots are presented in figures \ref{fig0}, \ref{fig1}and \ref{fig2}. In all the figures, $M=1$.

\clearpage

\section{Observables}\label{sec4}
A complete understanding of the shadow can be made by calculating $R_s$ which gives the approximate size of the shadow and $\delta_s$ which measures the deformation created in the shadow. To calculate these parameters we will follow \cite{Hioki}.
\noindent The silhouette of the shadow passes through three points. These are $T(\alpha_t,\beta_t)$ at the top, $B(\alpha_b,\beta_b)$ at the bottom,
$R(\alpha_r,0)$ at right. From these points, one can calculate $R_s$ and $\delta_s$. These reads

\begin{wrapfigure}[8]{l}{0.5\textwidth}
	\centering
	\includegraphics[width=0.5\textwidth]{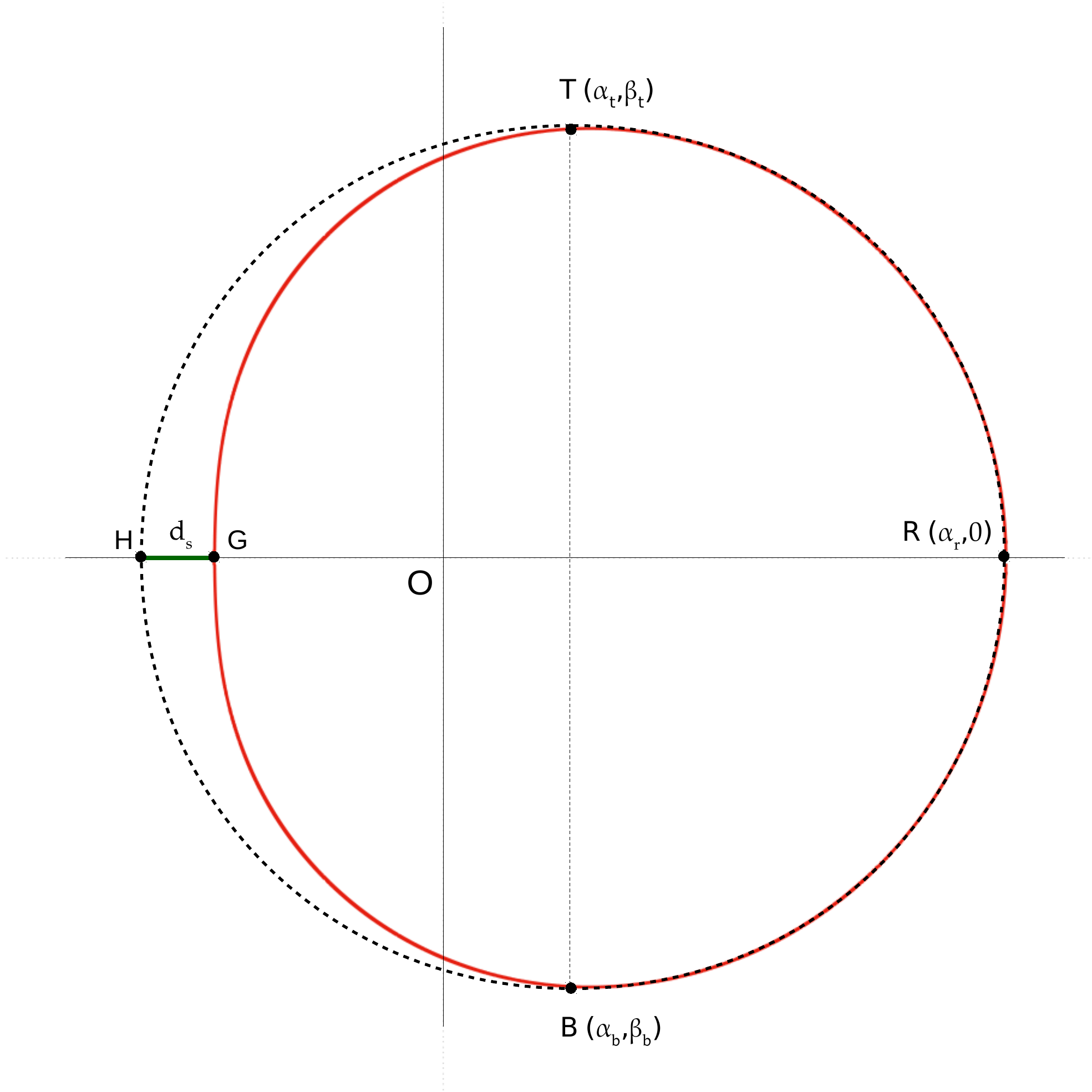}
	\caption{Representation of the shadow of a black hole.}
	\label{bs1}
\end{wrapfigure}
\begin{figure}[!h]
	\begin{minipage}[t]{0.48\textwidth}
		\centering\includegraphics[width=\textwidth]{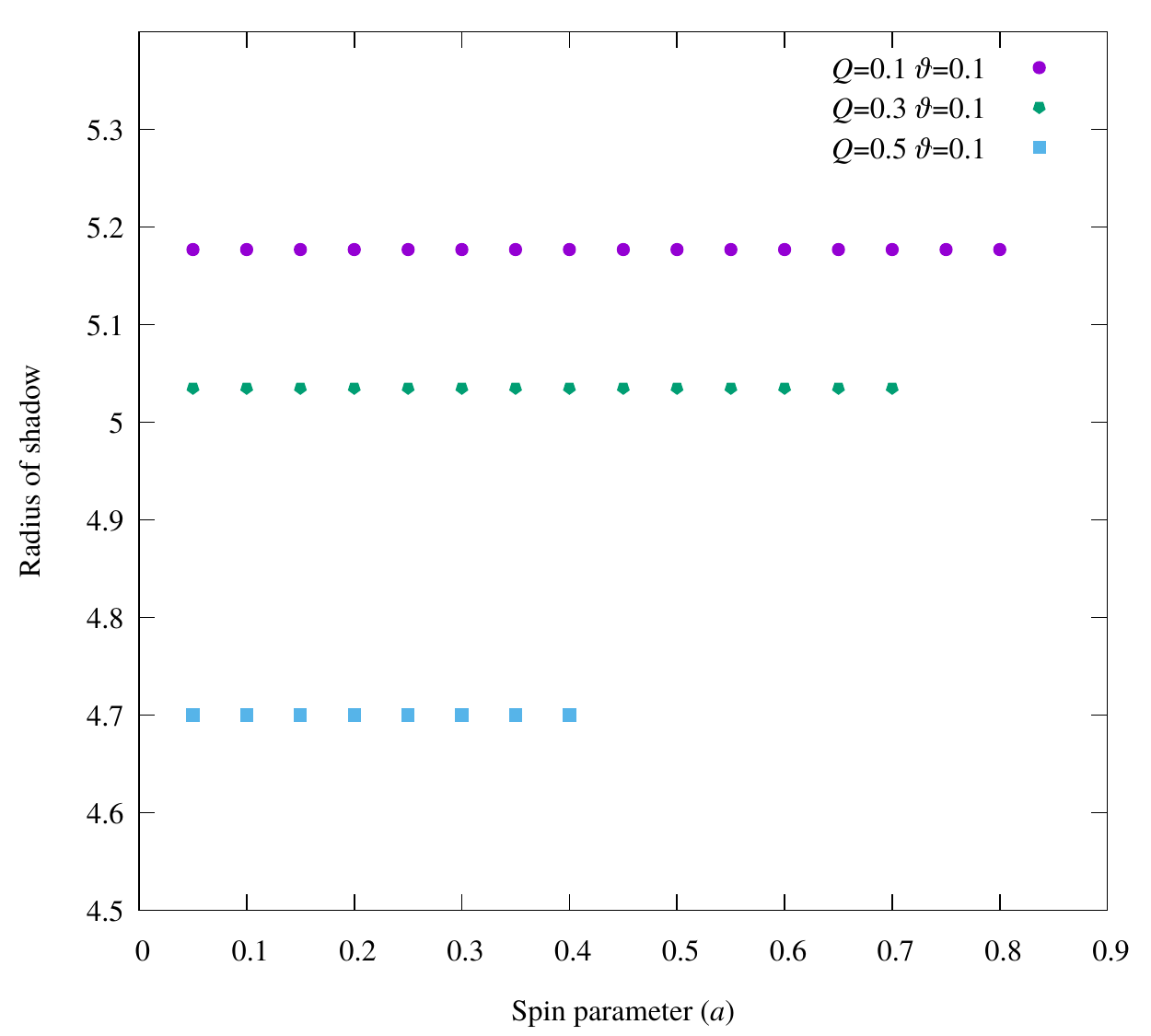}\\
		{\footnotesize Variation of Radius ($R_s$) of the shadows with Spin Parameter $a$}
	\end{minipage}\hfill
	\begin{minipage}[t]{0.48\textwidth}
		\centering\includegraphics[width=\textwidth]{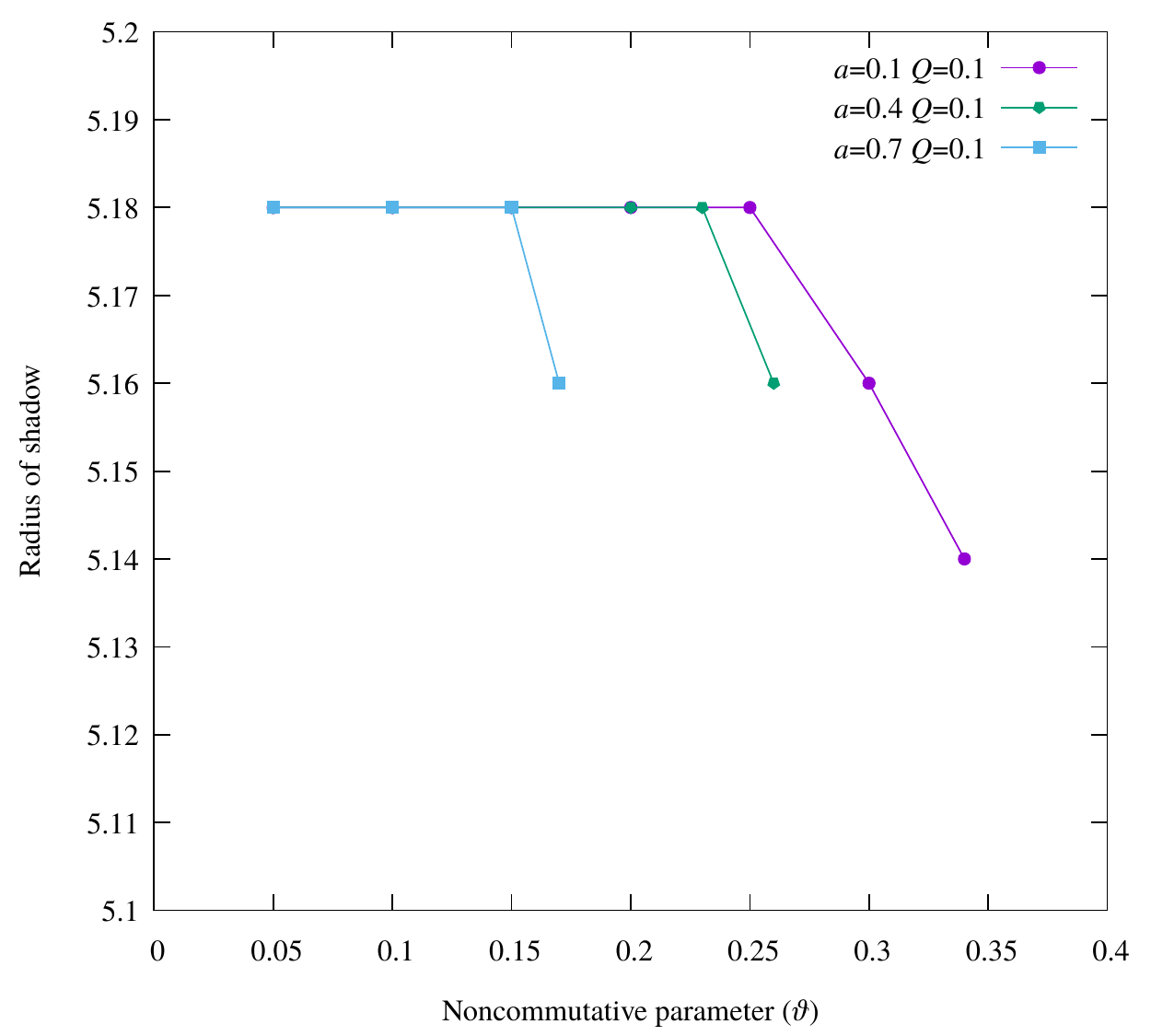}\\
		{\footnotesize Variation of Radius ($R_s$) of the shadows with noncommutative parameter $\vartheta$}
	\end{minipage}
	\caption{Comparison of variations of $R_s$}
	\label{fig9}
\end{figure}
\begin{figure}[!h]
	\begin{minipage}[t]{0.48\textwidth}
		\centering
		\includegraphics[width=\linewidth]{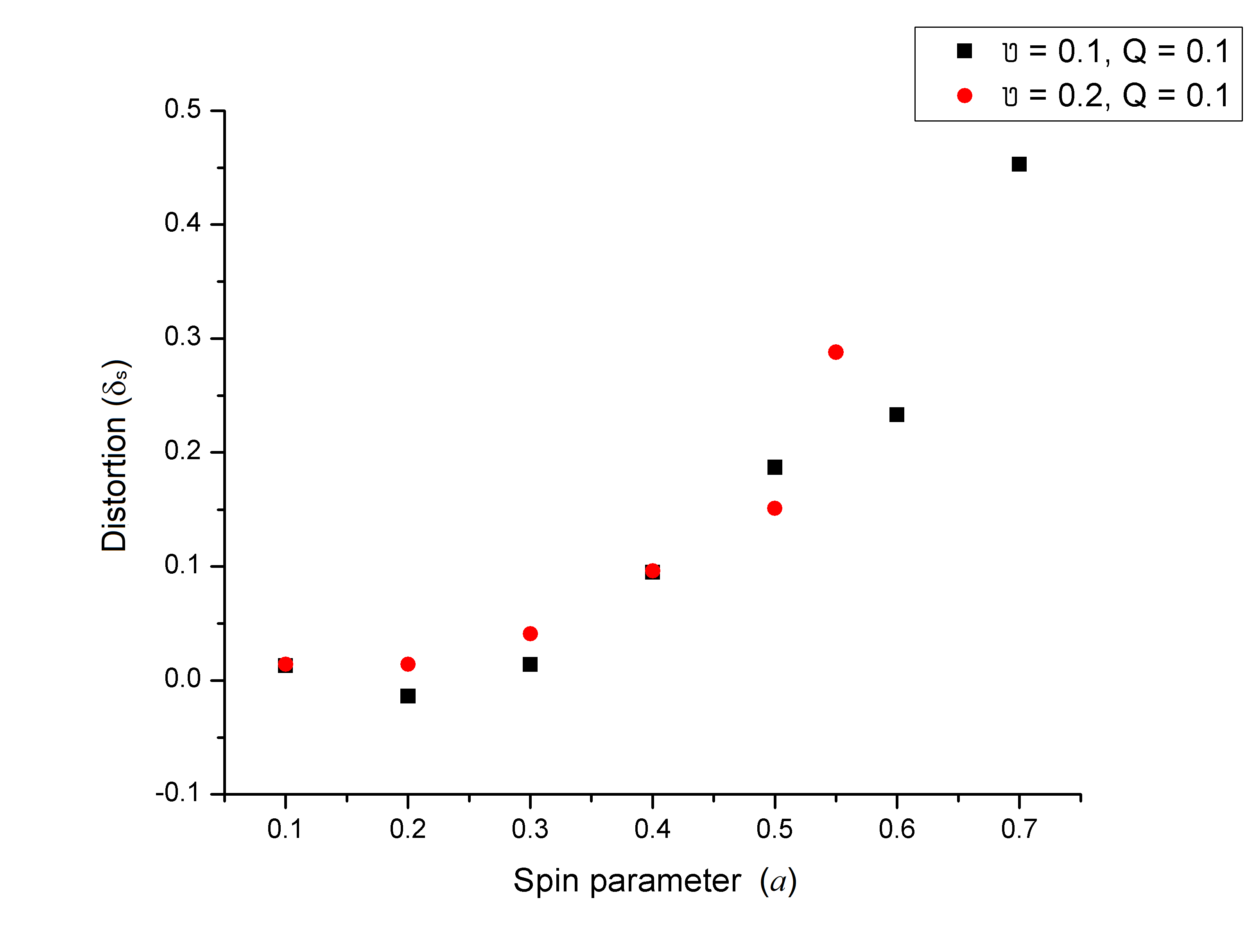}
		\caption{Variation of Distortion ($\delta_s)$ with $a$}
		\label{fig:ncrd}
	\end{minipage}\hfill
	\begin{minipage}[t]{0.48\textwidth}
		\centering
		\includegraphics[width=\linewidth]{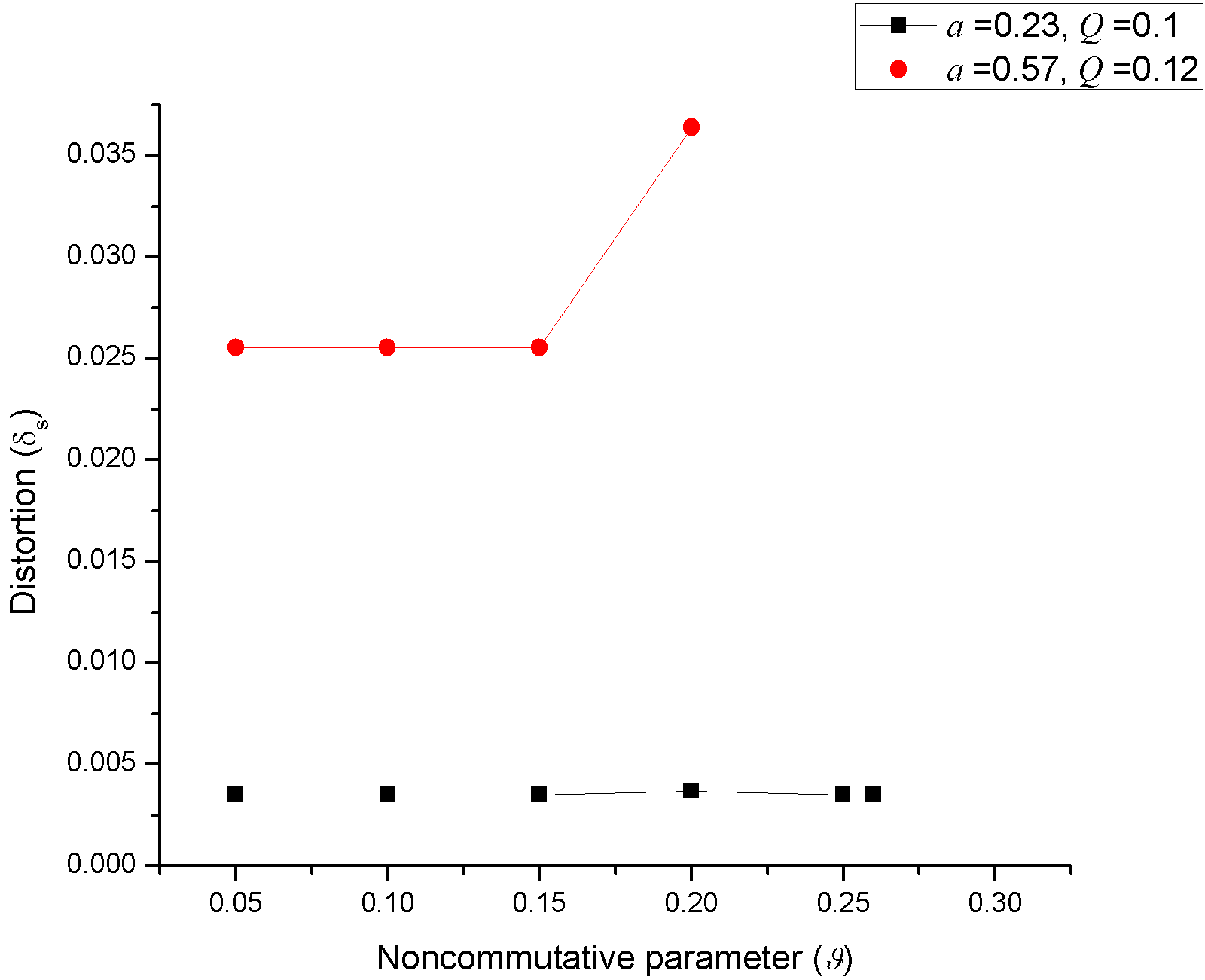}
		\caption{Variation of Distortion ($\delta_s)$ with $\vartheta$ }
		\label{fig:ncrdt}
	\end{minipage}\\
\end{figure}


\begin{eqnarray}
R_s=\frac{(\alpha_t-\alpha_r)^2+\beta_t^2}{2|\alpha_t-\alpha_r|}
\end{eqnarray}

\begin{eqnarray}
\delta_s= \frac{d_s}{R_s}=\frac{\bar{\alpha}_p-\alpha_p}{R_s}
\end{eqnarray}

\noindent where $H(\bar{\alpha}_p,0)$ and $G(\alpha_p,0)$ are the points arising due to the cut reference circle and shadow made to the horizontal axis at the left side of the center.
In figures \ref{fig9}, \ref{fig:ncrd} and \ref{fig:ncrdt}, we present the variations of $R_s$ and $\delta_s$ with the parameters $a$, $Q$ and $\vartheta$.

\clearpage

\section{Shadow in presence of plasma}\label{sec5}
In this section, we shall study the effects created by plasma on the shadow of the NC inspired ABG black hole. The refractive index of plasma is $n=n(x^i,\omega)$, $\omega$ is the photon frequency measured by the observer with velocity $u^\mu$. In this case the effective energy of photon is $\hbar\omega=-p_{\alpha}u^{\alpha}$.
The relation between plasma frequency and the 4-momentum of the photon reads \cite{Synge book}
\begin{eqnarray}
n^2=1+\frac{p_{\alpha}p^{\alpha}}{(p_{\mu}u^{\mu})^2}~.
\end{eqnarray}
It is expected that the presence of the plasma changes the shapes of the null-geodesics and this would in turn lead to a change in the celestial coordinates. The refractive index of plasma is connected to the plasma frequency as \cite{Rogers}
\begin{eqnarray}
n^2=1-\frac{\omega_p^2}{\omega^2}
\end{eqnarray}
where $\omega_p$ is the plasma frequency given by
$$\omega_p=\frac{4\pi e^2N(r)}{m_e}$$
where $e$ and $m_e$ represents the charge and mass of electron. $N(r)$ represents the number density of electrons in plasma. Following \cite{Synge book} and \cite{Rogers}, we take the variation of $N(r)$ with $r$ as the power-law, that is, $N(r)=\frac{N_0}{r^h}, h\ge0$. After assuming this form of $N(r)$, we can denote $\frac{\omega_p^2}{\omega^2}$ as
$$\frac{\omega_p^2}{\omega^2}=\frac{k}{r^h}$$
where $k\geq0$ is a constant and the case $k=0$ corresponds to the absence of plasma background. This in turn leads to
\begin{eqnarray}
n=\sqrt{1-\frac{k}{r^h}}~.
\end{eqnarray} 
Various values of $h$ can be taken which represents different physical properties about the nature of the plasma background. We will continue our calculation by taking $h=1$ so as to keep a minimum dependency on $r$ \cite{Rogers}.\\
\noindent The Hamilton-Jacobi equation in the plasma environment takes the form  \cite{Synge book}
\begin{eqnarray}
\frac{\partial S}{\partial \lambda}=-\frac{1}{2}\left[g^{\mu \nu}p_\mu p_\nu-(n^2-1)(p_0\sqrt{-g^{00}})^2\right]~.
\end{eqnarray}
These leads to the new set of null-geodesic equations
\begin{eqnarray}
r^2\dot{t} &=& a(L_z-an^2E)+\frac{r^2+a^2}{\Delta_{\vartheta}}\left[(r^2+a^2)n^2E-aL_z\right]\\[10pt]
r^2\dot{\phi} &=& (L_z-aE)+\frac{a}{\Delta_{\vartheta}}\left[E(r^2+a^2)-aL_z\right]\\[10pt]
r^2\dot{r} &=& \sqrt{\mathcal{R}_{pl}}\\[10pt]
r^2\dot{\epsilon} &=& \sqrt{\Theta_{pl}}
\end{eqnarray}
\noindent where 
\begin{eqnarray*}
\mathcal{R}_{pl} &=& \left[E(r^2+a^2)-aL_z\right]^2-\Delta_{\vartheta}\left[\mathcal{K}+(L_z-aE)^2\right]+(r^2+a^2)^2E^2(n^2-1)\\[7pt]
\Theta_{pl} &=& \mathcal{K}-(n^2-1)a^2E^2~.
\end{eqnarray*}
Hence, in the plasma environment, the effective potential for radial motion of photons reads
$$V_{eff}^{pl}=[E(r^2+a^2)-aL_z]^2-\Delta_{\vartheta}[\mathcal{K}+(L_z-aE)^2]+(r^2+a^2)^2E^2(n^2-1)~.$$
Proceeding as before, the conditions for unstable circular orbits now gives
\begin{eqnarray}
\left[(r^2+a^2)-a\xi\right]^2-\Delta_{\vartheta}\left[\eta+(\xi-a)^2\right]+(r^2+a^2)^2(n^2-1) &=& 0 \label{pl_1}\\[10pt]
4r\left[r^2+a^2-a\xi\right]+4r(n^2-1)(r^2+a^2)+2nn^\prime (r^2+a^2)^2-\Delta_{\vartheta}^\prime\left[\eta+(\xi-a)^2\right] &=& 0~. \label{pl_2}
\end{eqnarray}
Combining the above equations to solve for $\xi$ yields
$$A\xi^2-2B\xi+C=0$$
where
\begin{eqnarray*}
A=\frac{a^2}{\Delta_{\vartheta}}~,\quad B=\frac{a}{\Delta_{\vartheta}}\left(r^2+a^2-\frac{2\Delta_{\vartheta} r}{\Delta_{\vartheta}^\prime}\right)~,\\[10pt] C=\frac{n^2(r^2+a^2)}{\Delta_{\vartheta}}\left[r^2+a^2-\frac{2\Delta_{\vartheta}}{\Delta_{\vartheta}^\prime}\left(2r+\frac{n^\prime}{n}\right)\right].
\end{eqnarray*}
\noindent Solving the quadratic equation for $\xi$, we obtain
\begin{eqnarray}
\xi=\frac{B}{A}\pm \sqrt{\frac{B^2}{A^2}-\frac{C}{A}}~.
\label{xi}
\end{eqnarray}
Substituting eq.(\ref{xi}) in eq.(\ref{pl_2}) yields
\begin{eqnarray}
\eta=\frac{\left[r^2+a^2-a\xi\right]^2+(n^2-1)(r^2+a^2)^2}{\Delta_{\vartheta}}-\left(\xi-a\right)^2~.
\end{eqnarray}
The new set of values for $\frac{d\phi}{dr}$ and $\frac{d\epsilon}{dr}$ reads

\begin{eqnarray}
\frac{d\phi}{dr}&=&\frac{(L_z-aE)+ \frac{a}{\Delta_{\vartheta}}[(r_0^2+a^2)E-aL_z]}{\sqrt{[E(r_0^2+a^2)-aL_z]^2-\Delta_{\vartheta}[\mathcal{K}+(L_z-aE)^2]+(r_0^2+a^2)^2E^2(n^2-1)}}\\[10pt]
\frac{d\epsilon}{dr}&=&\frac{\sqrt{\mathcal{K}-(n^2-1)a^2E^2}}{\sqrt{[E(r_0^2+a^2)-aL_z]^2-\Delta_{\vartheta}[\mathcal{K}+(L_z-aE)^2]+(r_0^2+a^2)^2E^2(n^2-1)}}
\end{eqnarray}

\noindent which leads to
\begin{eqnarray}
\alpha=-\frac{\xi}{n}~, \qquad \beta=\frac{\sqrt{\eta+a^2-a^2n^2}}{n}~.
\end{eqnarray}

\begin{figure}[!h]
	\begin{minipage}[t]{0.48\textwidth}
		\centering\includegraphics[width=\textwidth]{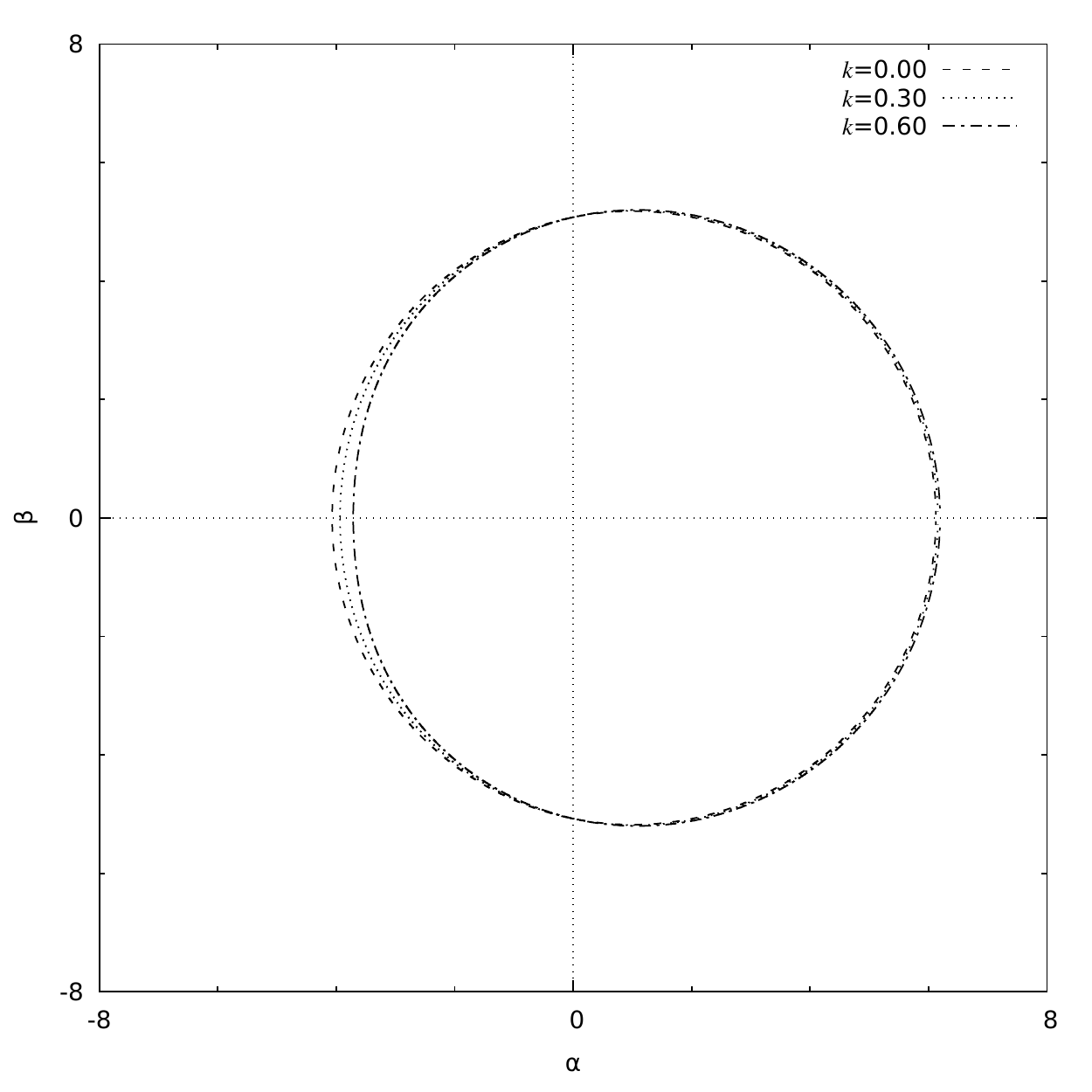}\\
		{\footnotesize $a=0.5$, $Q=0.1$, $\vartheta=0.1$.}
	\end{minipage}\hfill
	\begin{minipage}[t]{0.48\textwidth}
		\centering\includegraphics[width=\textwidth]{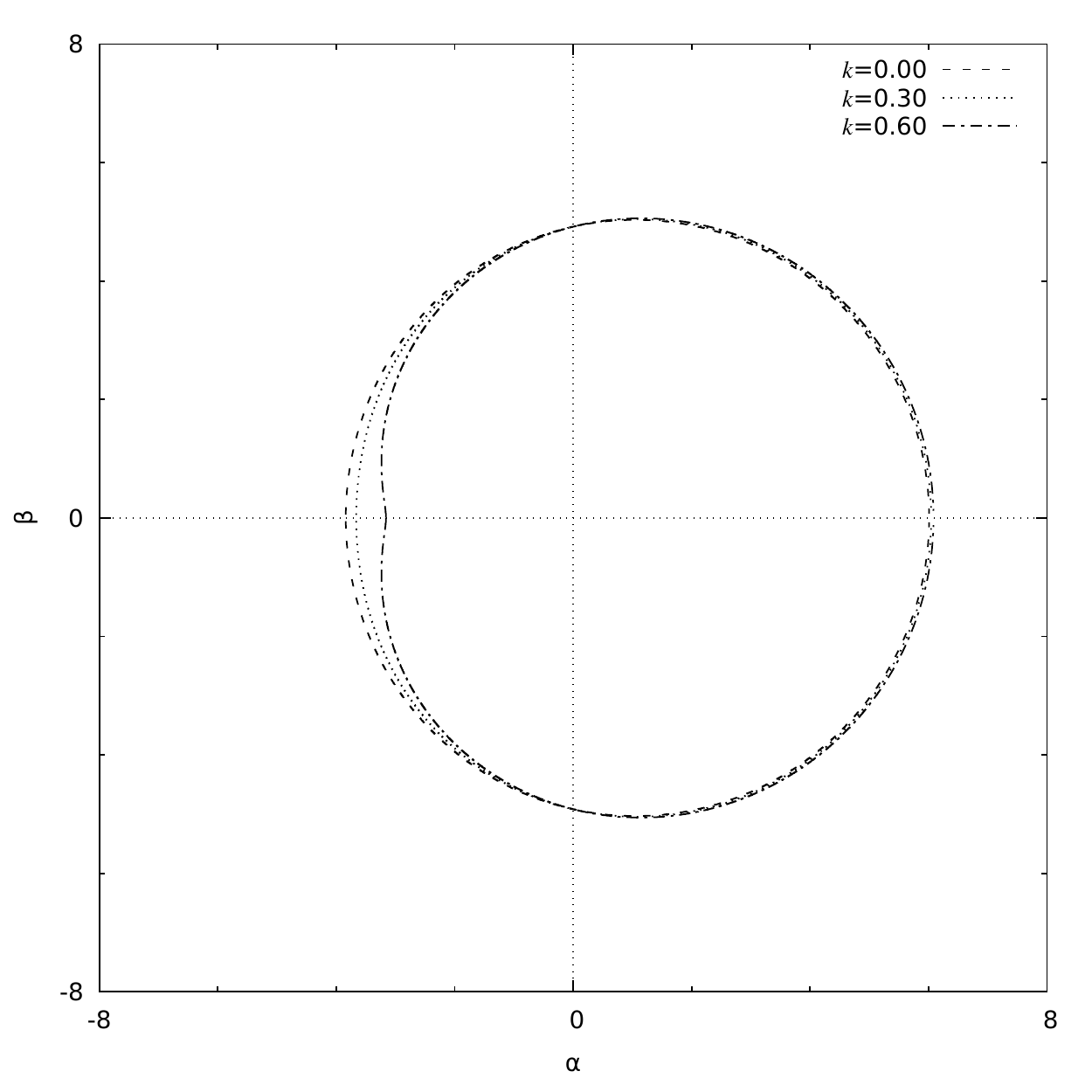}\\
		{\footnotesize $a=0.5$, $Q=0.3$, $\vartheta=0.1$.}
	\end{minipage}
	\caption{Shadow of the NC inspired ABG black hole in presence of a plasma background}
	\label{fig4}
\end{figure}

\begin{figure}[!h]
	\begin{minipage}[t]{0.48\textwidth}
		\centering\includegraphics[width=\textwidth]{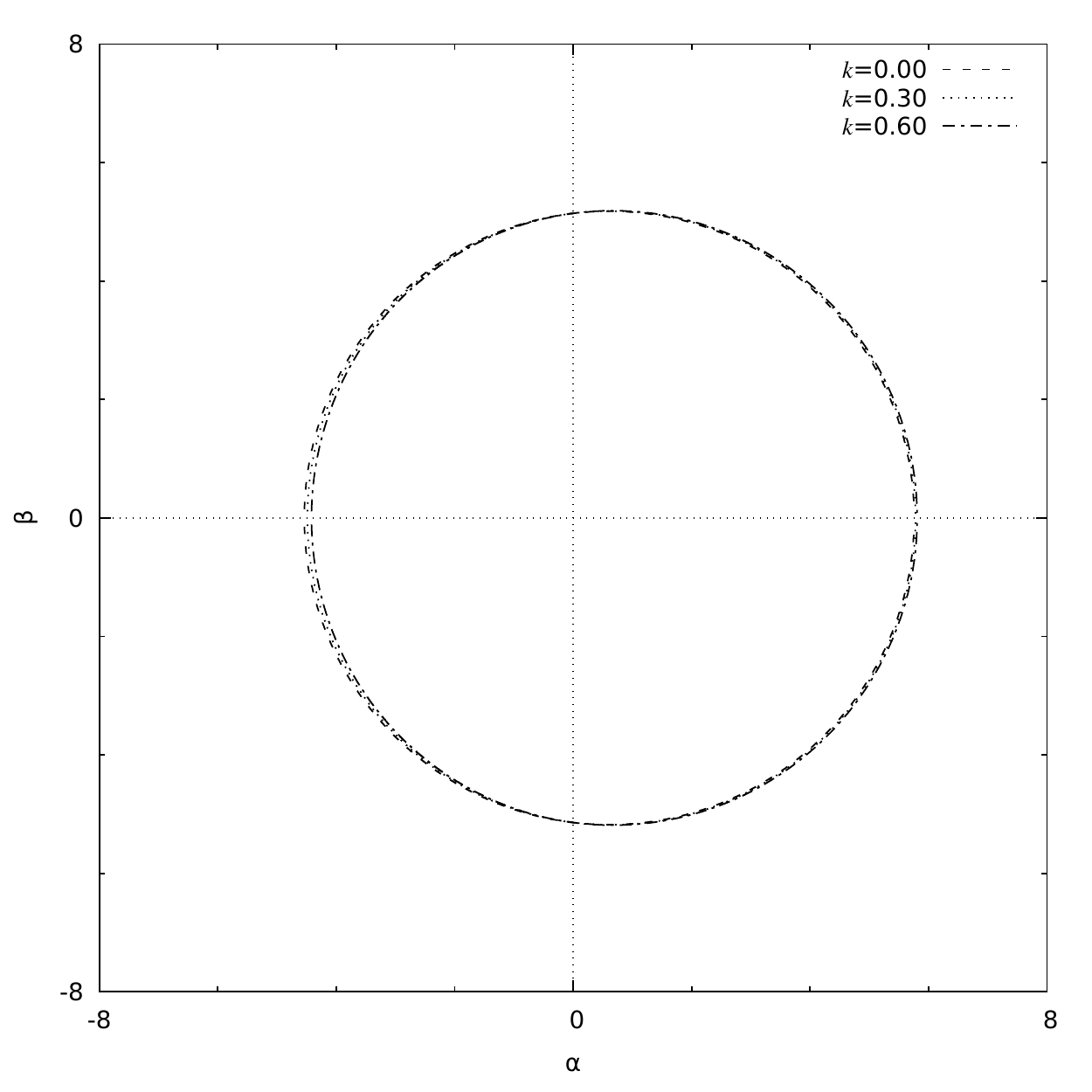}\\
		{\footnotesize $a=0.3$, $Q=0.1$, $\vartheta=0.2$.}
	\end{minipage}\hfill
	\begin{minipage}[t]{0.48\textwidth}
		\centering\includegraphics[width=\textwidth]{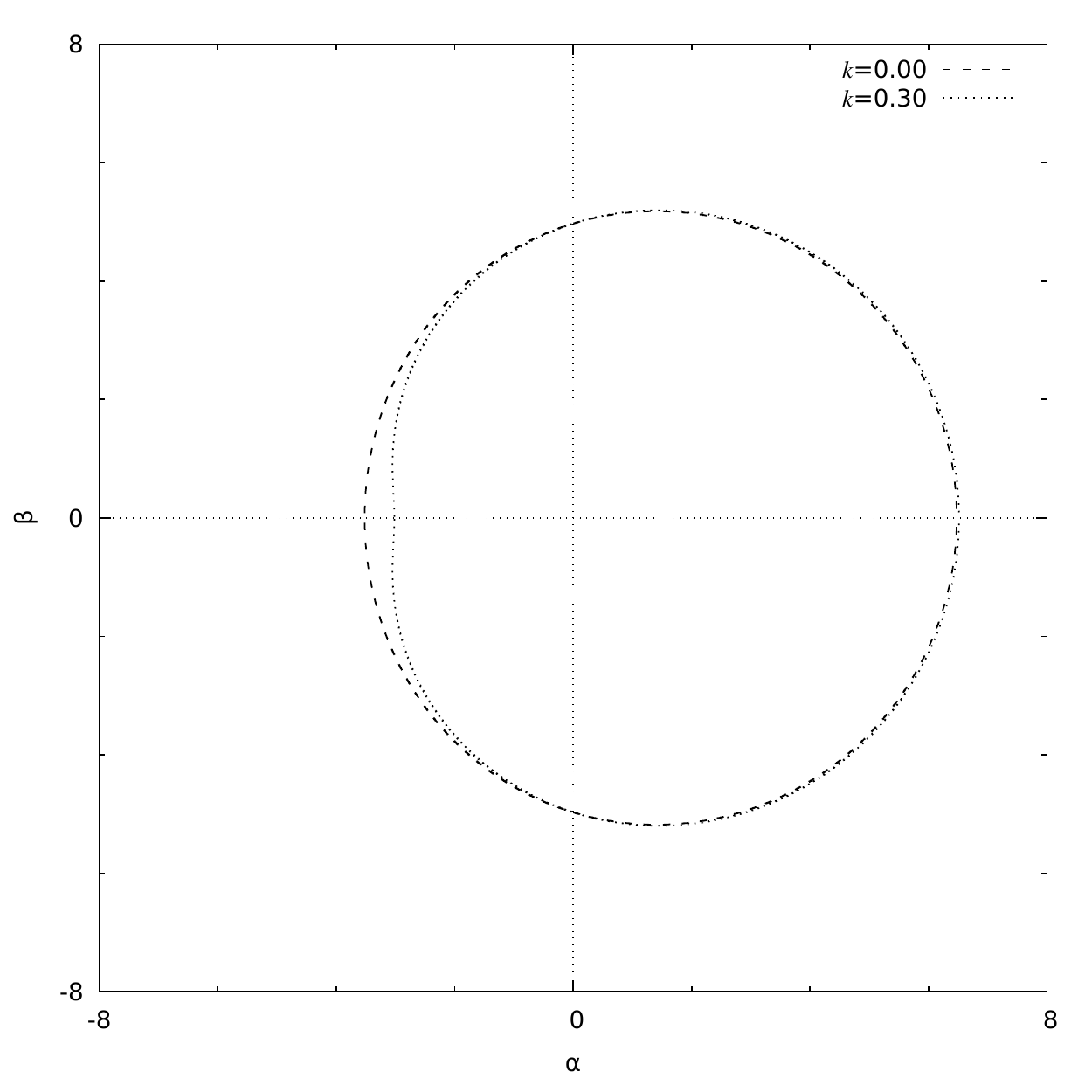}\\
		{\footnotesize $a=0.7$, $Q=0.1$, $\vartheta=0.1$.}
	\end{minipage}
	\caption{Shadow of the NC inspired ABG black hole in presence of a plasma background}
	\label{fig5}
\end{figure}

In figures \ref{fig4} and \ref{fig5}, we present the shadows of the NC inspired ABG black hole metric in the plasma background. In all the figures, $M=1$.

\clearpage

\begin{figure}[!h]
	\begin{minipage}[t]{0.48\textwidth}
		\centering\includegraphics[width=\textwidth]{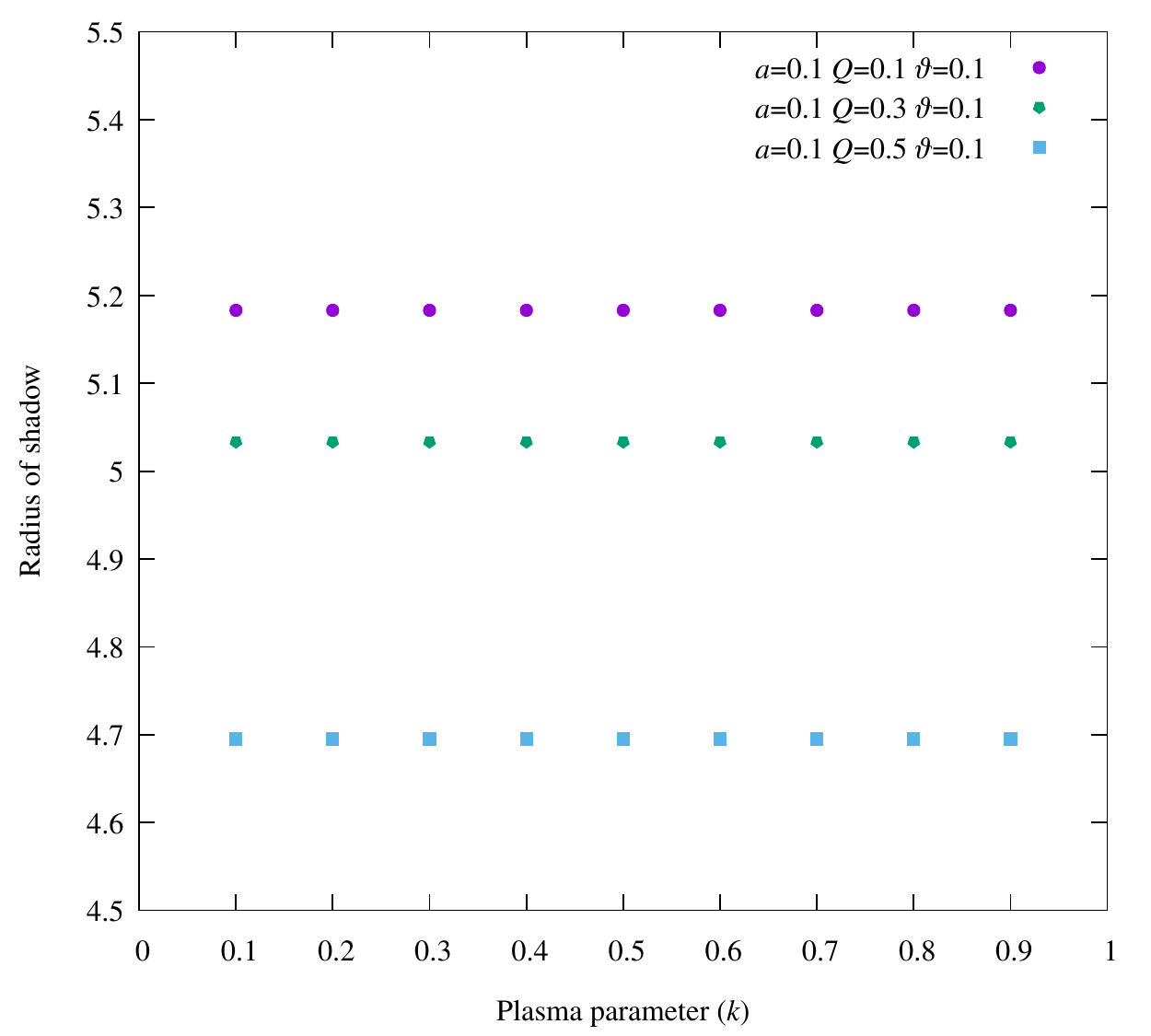}\\
	\end{minipage}\hfill
	\begin{minipage}[t]{0.48\textwidth}
		\centering\includegraphics[width=\textwidth]{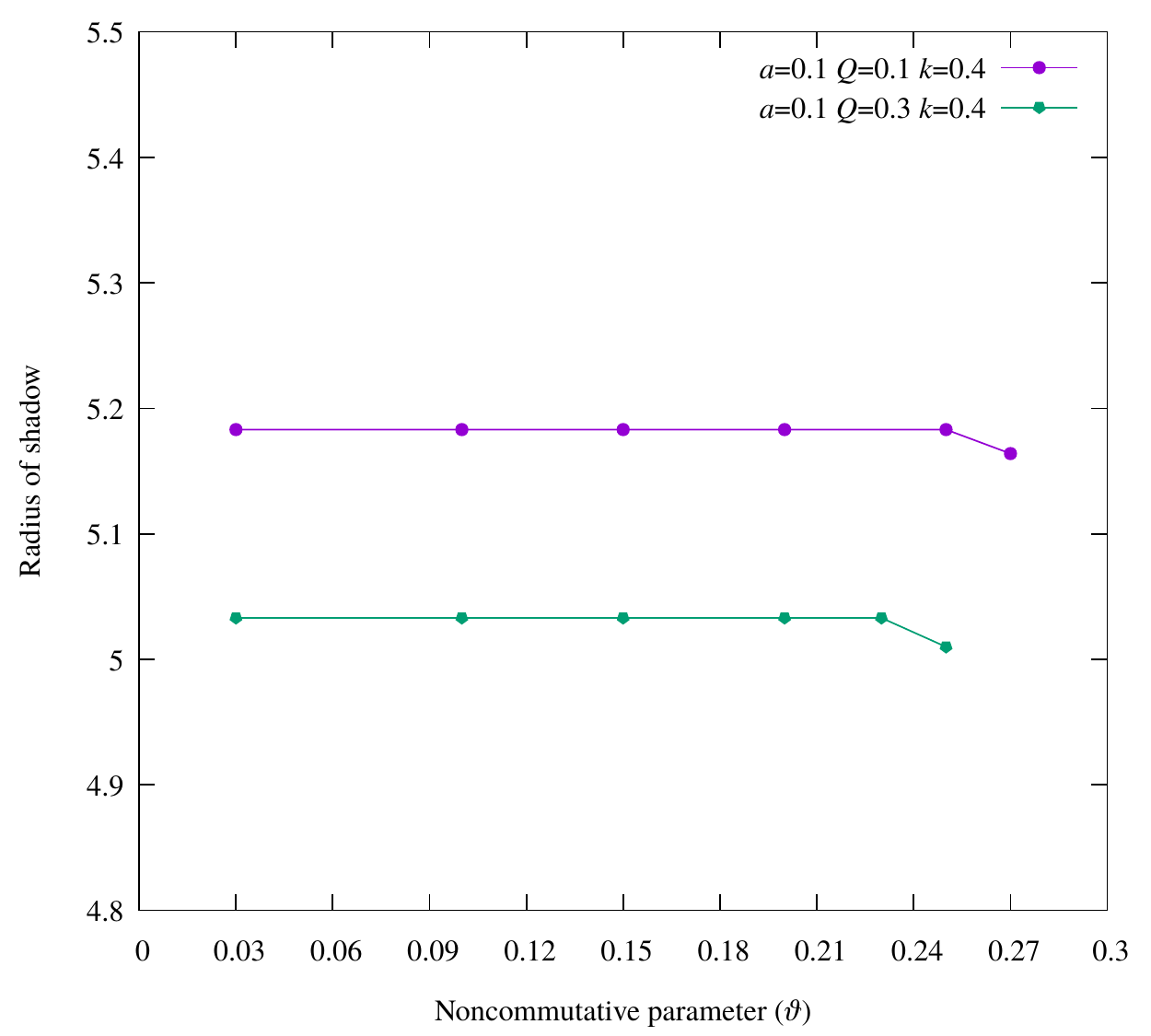}\\
	\end{minipage}
	\caption{Effect of plasma parameter $k$ and noncommutative parameter $\vartheta$ on radius of the shadow.}
	\label{fig6}
\end{figure}

\section{Conclusion}\label{sec6}

In this paper, we present the shadow of a noncommutative inspired Ay$\acute{o}$n Beato\@ Garc$\acute{i}$a (ABG) black hole, both in the presence and absence of plasma. We first introduce the noncommutative inspired ABG black hole metric and study its properties briefly. The effect of various parameters, namely, the spin parameter of the black hole $a$, the charge $Q$ and the noncommutative (NC) parameter $\vartheta$ on the event horizon is studied. We infer that for all values of spin parameter $a$ and charge $Q$, there exits a value of noncommutative parameter $\vartheta\ (\vartheta_{\text{max}})$ upto which the metric gives a black hole solution with at least one event horizon. We then construct the shadow of the NC inspired ABG black hole and plot it for different values of the spin parameter of the black hole $a$, the charge $Q$ and the noncommutative parameter $\vartheta$. We infer from the plots that a change in the spin parameter $a$ does not affect the size (radius $R_s$) of the shadow. However, an increase in the spin parameter $a$ increases the distortion ($\delta_s$) of the shadow. In contrast to the spin parameter, charge $Q$ and the NC parameter $\vartheta$ affect both the shape (distortion) and the size (radius) of the shadow. Higher values of charge $Q$ and the NC parameter $\vartheta$ make the distortion bigger when the spin parameter $a$ is high. Further, the noncommutative parameter $\vartheta$ does not affect the radius of the shadow if the value of $\vartheta$ is less than $\vartheta_{\text{max}}$, however, if $\vartheta > \vartheta_{\text{max}}$, the radius of the shadow ($R_s$) decreases. It should be noted that when $\vartheta > \vartheta_{max}$, the event horizon disappears and hence we do not have a black hole anymore. However, the shadow of the spacetime geometry still continues to exist up to some value of $\vartheta$ $(\vartheta > \vartheta_{\text{max}})$ which depends on the choice of the parameters $M$, $Q$ and $a$. We then introduce plasma into the NC inspired ABG metric and carry out our analysis. The results show that in the presence of the plasma background, the radius of the shadow does not change, but the distortion increases. We feel that these studies can in principle be relevant to detect the noncommutativity of spacetime.

\section{Acknowledgements}
AS acknowledges the support by Council for Scientific and Industrial Research (CSIR) for Junior Research Fellowship. SG acknowledges the support by DST-SERB under Start Up Research Grant (Young Scientist), File No. YSS/2014/000180. SG also acknowledges the support of IUCAA, Pune for the Visting Associateship Programme.

\end{document}

%% file: nc.tex
\begin{adjustwidth}{-0.25in}{-0.2in}
		\begin{minipage}[t][][t]{0.32\textwidth}
			\begin{tabular}{|c|c|c|}
				
				\hline
				$a\ \&\ Q$ & $\vartheta$ & $R_{EH}$\\ \hline\hline
				
				\multirow{2}{1em}{}$a$=0.88 & $\vartheta$=0.05 & 1.4337\\[2pt]
				$Q$=0.11 & $\vartheta_{max}=0.1$ & 1.3149 \\[3pt] \hline \hline
				
				\multirow{3}{4em}{} & $\vartheta$=0.05 & 1.4797\\
				{$a$}=0.86 & $\vartheta$=0.1 & 1.4268\\
				$Q$=0.1 & $\vartheta_{max}=0.11$ & 1.3487 \\[3pt] \hline \hline
				
				\multirow{3}{4em}{} & $\vartheta$=0.05 & 1.5247\\
				{$a$}=0.83 & $\vartheta$=0.1 & 1.4933\\
				$Q$=0.11 & $\vartheta_{max}=0.12$ & 1.3909 \\[3pt] \hline \hline
				
				\multirow{3}{4em}{} & $\vartheta$=0.05 & 1.5641\\
				{$a$}=0.8 & $\vartheta$=0.1 & 1.5432\\
				$Q$=0.12 & $\vartheta_{max}=0.13$ & 1.4148 \\[3pt] \hline \hline
				
				\multirow{3}{4em}{} & $\vartheta$=0.05 & 1.6025\\
				{$a$}=0.78 & $\vartheta$=0.1 & 1.5879\\
				$Q$=0.1 & $\vartheta_{max}=0.14$ &  1.4340\\[3pt] \hline\hline
				
				\multirow{3}{4em}{} & $\vartheta$=0.05 & 1.6397\\
				{$a$}=0.75 & $\vartheta$=0.1 & 1.6295\\
				$Q$=0.1 & $\vartheta_{max}=0.15$ & 1.4600 \\[3pt] \hline\hline
				
				\multirow{4}{4em}{} & $\vartheta$=0.05 & 1.6737\\
				{$a$}=0.72 & $\vartheta$=0.1 & 1.6661\\
				& $\vartheta$=0.15 & 1.5693\\
				$Q$=0.1 & $\vartheta_{max}=0.16$ & 1.4711 \\[3pt] \hline\hline
				
				\multirow{4}{4em}{} & $\vartheta$=0.05 & 1.7102\\
				{$a$}=0.68 & $\vartheta$=0.1 & 1.7049\\
				& $\vartheta$=0.15 & 1.6387\\
				$Q$=0.11 & $\vartheta_{max}=0.17$ & 1.5142 \\ \hline
				
			\end{tabular}
		\end{minipage}\hfill
		\begin{minipage}[t][][t]{0.32\textwidth}
			\begin{tabular}{|c|c|c|}
				
				\hline
				$a\ \&\ Q$ & $\vartheta$ & $R_{EH}$\\ \hline \hline
				
				\multirow{4}{4em}{} & $\vartheta$=0.05 & 1.7418\\
				{$a$}=0.65& $\vartheta$=0.1 & 1.7378\\
				& $\vartheta$=0.15 & 1.6873\\
				$Q$=0.1 & $\vartheta_{max}=0.18$ & 1.5252 \\ \hline\hline
				
				\multirow{4}{4em}{} & $\vartheta$=0.05 & 1.7716\\
				{$a$}=0.61& $\vartheta$=0.1 & 1.7686\\
				& $\vartheta$=0.15 & 1.7289\\
				$Q$=0.11 & $\vartheta_{max}=0.19$ & 1.5363 \\ \hline\hline
				
				\multirow{4}{4em}{} & $\vartheta$=0.05 & 1.7980\\
				{$a$}=0.57& $\vartheta$=0.1 & 1.7957\\
				& $\vartheta$=0.15 & 1.7633\\
				$Q$=0.12 & $\vartheta_{max}=0.20$ & 1.5079 \\ \hline\hline
				
				\multirow{5}{4em}{} & $\vartheta$=0.05 & 1.8290\\
				{$a$}=0.53 & $\vartheta$=0.1 & 1.8272\\
				& $\vartheta$=0.15 & 1.8013\\
				& $\vartheta$=0.20 & 1.6683\\
				$Q$=0.11 & $\vartheta_{max}=0.21$ & 1.5503 \\ \hline\hline
				
				\multirow{5}{4em}{} & $\vartheta$=0.05 & 1.8556\\
				{$a$}=0.48 & $\vartheta$=0.1 & 1.8542\\
				& $\vartheta$=0.15 & 1.8328\\
				& $\vartheta$=0.20 & 1.7326\\
				$Q$=0.12 & $\vartheta_{max}=0.22$ & 1.5539 \\ \hline\hline
				
				\multirow{5}{4em}{} & $\vartheta$=0.05 & 1.8853\\
				{$a$}=0.43 & $\vartheta$=0.1 & 1.8843\\
				& $\vartheta$=0.15 & 1.8669\\
				& $\vartheta$=0.20 & 1.7883\\
				$Q$=0.11 & $\vartheta_{max}=0.23$ & 1.5943 \\ \hline
				
			\end{tabular}
		\end{minipage}\hfill
		\begin{minipage}[t][][t]{0.32\textwidth}
			\begin{tabular}{|c|c|c|}
				\hline
				$a\ \&\ Q$ & $\vartheta$ & $R_{EH}$\\ \hline \hline
				
				\multirow{5}{4em}{} & $\vartheta$=0.05 & 1.9122\\
				{$a$}=0.37 & $\vartheta$=0.1 & 1.9114\\
				& $\vartheta$=0.15 & 1.8969\\
				& $\vartheta$=0.20 & 1.8325\\
				$Q$=0.11 & $\vartheta_{max}=0.24$ & 1.6147 \\[5pt] \hline\hline
				
				\multirow{5}{4em}{} & $\vartheta$=0.05 & 1.9372\\
				{$a$}=0.31 & $\vartheta$=0.1 & 1.9366\\
				& $\vartheta$=0.15 & 1.9244\\
				& $\vartheta$=0.20 & 1.8700\\
				$Q$=0.1 & $\vartheta_{max}=0.25$ & 1.6180 \\[5pt] \hline\hline
				
				\multirow{6}{4em}{} & $\vartheta$=0.05 & 1.9601\\
				{$a$}=0.23 & $\vartheta$=0.1 & 1.9596\\
				& $\vartheta$=0.15 & 1.9492\\
				& $\vartheta$=0.20 & 1.9023\\
				$Q$=0.1 & $\vartheta$=0.25 & 1.7392\\
				& $\vartheta_{max}=0.26$ & 1.6067 \\[5pt] \hline\hline
				
				\multirow{6}{4em}{} & $\vartheta$=0.05 & 1.9999\\
				{$a$}=0 & $\vartheta$=0.1 & 1.9996\\
				& $\vartheta$=0.15 & 1.9916\\
				& $\vartheta$=0.20 & 1.9543\\
				$Q$=0 & $\vartheta$=0.25 & 1.8418\\
				& $\vartheta_{max}=0.275$ & 1.6395 \\[5pt] \hline

			\end{tabular}
		\end{minipage}
\end{adjustwidth}

%% file: paper.bbl
\begin{thebibliography}{99}
	
	\bibitem{Synge}J.\@ L.\@ Synge, Mon. Not. R. Astron. Soc. 131, 463 (1966).
	\bibitem{Bardeen}J.\@ M.\@ Bardeen, in ``Black Holes" (Les Astres Occulus), edited by C.\@ Dewitt and B.\@ S.\@ Dewitt (Gordon and Breach, 1973), pp. 215-239.
	\bibitem{Kerr-Newmann Shadow}A.\@ de.\@ Vries, Class. Quantum Grav. 17 123 (2000)
	\bibitem{Hioki}K.\@ Hioki and K.\@ I.\@ Maeda,Phys. Rev. D 80, 024042 (2009).
	\bibitem{Ghosh}M.\@ Amir, S.\@ G.\@ Ghosh, Phys. Rev. D 94, 024054 (2016).
	\bibitem{5DBH}U.\@ Papnoi, F.\@ Atamurotov, S.\@ G.\@ Ghosh, B.\@ Ahmedov, Phys. Rev. D 90, 024073 (2014).
	\bibitem{EHT}The Event Horizon Telescope - www.eventhorizontelescope.org.
	\bibitem{Nicolini}P.\@ Nicolini, Int. J. Mod. Phys. A 24,1229 (2009).
	\bibitem{SG}R.\@ Banerjee, S.\@ Gangopadhyay, S.\@ K.\@ Modak, Phys. Lett. B 686, 181-187 (2010).
	\bibitem{SG 2}S.\@  Gangopadhyay, Journal of Physics: Conference Series 405 (2012) 012014.
	\bibitem{Maggiore}M.\@ Maggiore, Phys. Lett. B 304, 65 (1993).
	\bibitem{epjc}M.\@ Sharif, S.\@ Iftikar, Eur. Phys. J. C 76, 630 (2016).
	\bibitem{ABG}E.\@ Ay$\acute{o}$n-Beato and A.\@ Garc$\acute{i}$a, Phys. Rev. Lett. 80, 5056 (1998).
	\bibitem{rotating abg}B.\@ Toshmatov, B.\@ Ahmedov, A.\@ Abdujabbarov, Z.\@ Stuchlik, Phys. Rev. D 89, 104017 (2014).
	\bibitem{rotating abg 2}M.\@ Azreg-A$\ddot{1}$nou, Phys. Rev. D 90, 064041 (2014).
	\bibitem{newman-janis}Drake and Szekeres, General Relativity and Gravitation (2000) 32: 445.
	\bibitem{E spallucci}E.\@ Spallucci, A.\@ Smailagic, P.\@ Nicolini, Phys. Lett. B 670, 449 (2009).
	\bibitem{prd}A.\@ Ahmadjon, M.\@ Amir, B.\@ Ahmedov and S.\@ G.\@ Ghosh, Phys. Rev. D 93, 104004 (2016).
	\bibitem{Derek}D.\@ Raine, E.\@ Thomas, ``Black Holes: an introduction" $2^{nd}$ ed. Imperial College Press.
	\bibitem{carter}B.\@ Carter, Phys. Rev. 174, 1559 (1968).
	\bibitem{chandrashekhar}S.\@ Chandrasekhar,``The Mathematical Theory of Black Holes", Oxford University Press, 1998.
	\bibitem{celes.cord.}A.\@ E.\@ Vazquez, E.\@ P.\@ Esteban, Nuovo Cim. B 119, 489 (2004).
	\bibitem{Synge book}J.\@ L.\@ Synge,``Relativity: The General Theory", North Holland, Amsterdam, 1960.
	\bibitem{Rogers}A.\@ Rogers, Mon. Not. R. Astron. Soc. 451, 4536 (2015).
	
\end{thebibliography}
